\documentclass[aps,twocolumn,prb]{revtex4}
\usepackage{subfigure}
\usepackage{amssymb}
\usepackage{amsmath}
\usepackage{amsfonts}
\usepackage{latexsym}
\usepackage{graphicx}
\usepackage[usenames,dvipsnames]{color}
\usepackage[all]{xy}
\usepackage[normalem]{ulem}

\usepackage{longtable}

\begin{document}
\newcommand{\av}[1]{{\color{blue}$\clubsuit$#1}}
\newcommand{\cor}[1]{{\color{red}$\spadesuit$#1}}
\newcommand{\sg}[1]{(SG\##1)}
\newcommand{\cm}[1]{\sout{#1}}

\title{Stability of 41 metal - boron systems at 0 GPa and 30 GPa from first principles}
\author{A.G. Van Der Geest$^{\ddag}$, A.N.  Kolmogorov$^\ddag$}
\date{\today}
\affiliation{$^\ddag$ State University of New York - Binghamton, USA\\Corresponding author: A.N. Kolmogorov $<$kolmogorov@binghamton.edu$>$}

\begin{abstract}
\noindent{
A multitude of observed boron-based materials have outstanding superconducting, mechanical, and refractory properties. Yet, the structure, the composition, and the very existence of some reported metal boride (M-B) compounds have been a subject of extensive debate. This density functional theory work seeks to set a baseline for current understanding of known metal boride phases as well as to identify new synthesizable candidates. We have generated a database of over 12,000 binary M-B entries for pressures of 0 and 30 GPa producing the largest scan of compositions and systems in this materials class. The 175 selected crystal structures include both observed prototypes and new ones found with our evolutionary ground state search. The metals considered are:  Al, Ag, Au, Ba, Be, Ca, Cd, Co, Cr, Cs, Cu, Fe, Hf, Hg, Ir, K, La, Li, Mg, Mn, Mo, Na, Nb, Ni, Os, Pd, Pt, Rb, Re, Rh, Ru, Sc, Sr, Ta, Tc, Ti, V, W, Y, Zn, and Zr. Based on the formation energy calculated at zero pressure and temperature 4 new M-B phases or structures have been predicted, while a number of previously reported compounds have been shown to be unstable. At 30 GPa, changes in the convex hulls are expected to occur in 18 out of 41 M-B systems, which is used to indicate regions of the periodic table (for metal borides) that require further investigation from the community.  Analysis of the collected information has revealed a nearly linear relationship between the magnetic moment per atom and the metal content for all the Fe-B, Co-B, and Ni-B structures within 0.15 eV/atom of the stability tie line. Both GGA-PBE and LDA-PW functionals were used to provide an understanding of the systematic error introduced by the choice of the exchange-correlation functional.}
 \end{abstract}

\maketitle

\section{Introduction}

Boron forms compounds with nearly all metals and defines a remarkable variety of morphologies in compounds, including 3D polyhedron frameworks, 2D nets, and 1D chains~\cite{ Albert2000b,Etourneau2007}. The richness and the complexity of metal boride structures can be traced back to the behavior of pure boron. Namely, the tendency of the second-row element to form covalent bonds sets boron apart from the neighboring nearly-free-electron metals in the periodic table, such as magnesium or aluminum~\cite{pettifor1995}. At the same time, the three valence electrons are insufficient to fill up all the bonding states in stable covalent structures adopted by tetravalent elements, such as carbon or silicon~\cite{pettifor1995}. The resulting frustration makes boron a metalloid with arguably the most complex non-magnetic elemental ground state under normal conditions~\cite{Parakhonskiy2011,Kurakevych2012,Albert2009}. For example, the stability of the known ambient-pressure $\alpha$-B and $\beta$-B polymorphs along with the recently discovered high-pressure $\gamma$-B polymorph has been subject of numerous recent studies~\cite{Parakhonskiy2011, Oganov2009, Kurakevych2012, Albert2009, Zarechnaya2008}. 

Not surprisingly, mixing boron with metals of different size and valence has created a large class of inorganic compounds with diverse mechanical and electronic properties, some of which have been uncovered only in recent years~\cite{massalski1990, Nagamatsu2001, Arima2013, Bose2005, Fang2009, Gou2013}. The most striking example is the 2001 discovery of a phonon-mediated superconducting transition in a well-known binary compound, MgB$_2$, at a record-breaking 39 K~\cite{Nagamatsu2001}. The research generated by this breakthrough has identified unusual non-centrosymmetric superconductors (Li$_2$(Pd,Pt)$_3$B ~\cite{Arima2013, Bose2005, Togano2004, Badica2004} and Ru$_7$B$_3$~\cite{Fang2009}) and has led to a number of predicted conventional superconducting boron-based materials (Li$_x$BC~\cite{Rosner2002}, LiB~\cite{Kolmogorov2006b,Kolmogorov2008}, Li-TM-B~\cite{Kolmogorov2008}, and  Fe-B~\cite{Kolmogorov2010}).  One of the proposed new materials, FeB$_4$, has been synthesized recently  and appears to be the first superconductor designed entirely on the computer~\cite{Gou2013}. Due to the   super hardness of pure boron\cite{Oganov2009b}, metal borides have also attracted attention as candidate cheap hard materials~\cite{Levine2009}.  ReB$_2$~\cite{Chung2007} and CrB$_4$~\cite{Niu2012} have been suggested to have unusually high hardness, but further investigations have indicated that they are not super hard~\cite{Dubrovinskaia2007, Qin2008, Knappschneider2013}. A lot of recent work has been dedicated to related transition-metal borides, such as WB$_3$/WB$_4$~\cite{Zhang2012, Chen2008, Liang2011, Liang2012, Liang2013} and the Mo-B system~\cite{Shein2007, Zhang2010, Liang2012b}. The chemical inertness and low volatility of metal borides have also allowed for their commercial use as refractory materials~\cite{Lundstrom1985}, thermionic emitters~\cite{Ji2011,Yada1989}, and steel strengthening agents~\cite{Lanier1994}.   It is worth noting that the strongest permanent magnets include a small amount of boron in a complex Nd-Dy-Fe-B compound ~\cite{Gutfleisch2011}.

Over the last decades the techniques used to investigate metal boride structures have varied greatly. Leading up to and during the 1960's a majority of the known metal borides were synthesized for the first time~\cite{Aronsson1968, Aronsson1965,Matkovic1965}. Development of chemically intuitive semi-empirical electronic structure methods during the 1970's and 1980's has been instrumental for building structural and binding models for selected compounds. Examples include the studies of the La and Y hexaborides~\cite{Walch1977}, Mn and Cr tetraborides \cite{Burdett1988}, and several other Cr borides~\cite{Okada1987}. Advances in density functional theory (DFT)~\cite{Perdew1996,perdew1981,Kresse1999} during the 1990's into the 2000's enabled characterization of materials through systematic calculations of their stability across the periodic table.  For example, \citet{Oguchi2002} studied binding trends for the AlB$_2$ prototype, while \citet{Kolmogorov2006} rationalized the stability of several metal boride structures. 

Screening large libraries of \emph{known} crystal structures and various compositions has emerged as an increasingly popular approach to not only identify new materials, but to also check the reliability of DFT-based methods~\cite{Curtarolo2005a, Levy2010, Levy2010b, Jain2011, Jain2011b, Bil2011}. \citet{Curtarolo2005a} showed that formation energy evaluated with DFT at zero temperature is a reliable criterion for compound existence giving a 92.4\% agreement (97\% with experimental error removed) with experiment for 80 binary metal systems. A number of computational frameworks have been developed, such as AFLOW \cite{Curtarolo2012} or the Materials Project~\cite{Jain2011}, that enable high-throughput screening of available inorganic databases for new materials with targeted properties \cite{Curtarolo2013}. 

Other search techniques seek to identify new stable materials crystallizing in yet unknown configurations. The problem of global crystal structure optimization (Ôfrom scratchÕ) has been addressed with the introduction of such approaches as evolutionary algorithms ~\cite{Bush1995, Abraham2006, oganov2006,Trimarchi2007}, minima hopping ~\cite{Wales1997, Wales1999, Amsler2010}, random searches ~\cite{Pickard2006}, or particle swarm optimizations~\cite{Wang2010,Wang2012}. Ground state searches for large systems are typically carried out at fixed compositions due to high computational cost, although variable-composition optimization has also been proposed~\cite{Trimarchi2009}.  \citet{Meredig2013} have developed techniques, called FPASS, to combine the experimentally known properties of structures such as space group and lattice constant with evolutionary algorithms to determine the atomic structure of unrefined powder diffraction files. Synthesis and characterization of complex high-pressure phases of B \cite{Lyakhov2010, Lyakhov2013} and CaB$_6$~\cite{Kolmogorov2012} have illustrated (i) the advantages of the evolutionary approach to constructing large ground states Ôfrom scratchÕ and (ii) the considerable benefits of incorporating prior structural information into the searches. Complementary ground state searches at selected compositions beyond the known prototypes have been performed in several detailed studies of binary M-B systems. Some of the predictions have already been confirmed, such as the incompleteness of the previous structural model of the known CrB$_4$ compound\cite{Kolmogorov2010,bialon2011,Niu2012} and the existence of a new FeB$_4$ material (meta)stable under normal conditions\cite{Kolmogorov2010,bialon2011,Gou2013}, while others are awaiting confirmation, such as the possibility of new Li-B~\cite{Kolmogorov2006b,Hermann2013b,Peng2012}, Fe-B~\cite{Kolmogorov2010,bialon2011}, Mo-B~\cite{Zhang2010}, W-B~\cite{Liang2013}, and Ca-B\cite{Kolmogorov2012,shah2013} phases under ambient or high pressures.

The next natural step in the investigation of this class of materials is a systematic scan of all relevant metal-boron binary systems with a combination of high-throughput and evolutionary-type calculations. The generation of the largest \emph{ab initio} database for metal borides has given us an opportunity to (i) identify new stable candidate materials for further investigation; (ii) establish trends across M-B systems and compositions; and (iii) expand the series of benchmark studies~\cite{Curtarolo2005a, Levy2010, Levy2010b, Jain2011, Jain2011b, Bil2011} that reveal the current capabilities of DFT-based methods for guiding materials research. Known metal boride structures found in the Inorganic Crystal Structure Database (ICSD)~\cite{ICSD1987} were considered along with new structures generated from our evolutionary search. The studied set of $s$-$p$ and transition metals is: Al, Ag, Au, Ba, Be, Ca, Cd, Co, Cr, Cs, Cu, Fe, Hf, Hg, Ir, K, La, Li, Mg, Mn, Mo, Na, Nb, Ni, Os, Pd, Pt, Rb, Re, Rh, Ru, Sc, Sr, Ta, Tc, Ti, V, W, Y, Zn, and Zr. We carried out calculations at both ambient (0 GPa) and high (30 GPa) pressure. The value of 30 GPa was chosen to bracket the pressure range of up to $\sim$25 GPa in which facile materials synthesis can be carried out with multi-anvil cells. In addition, materials with a certain degree of covalent bonding synthesized under such moderate pressures have been known to remain metastable once quenched down to normal conditions ($\gamma$-B~\cite{Oganov2009}, CaB$_6$~\cite{Kolmogorov2012}, FeB$_4$\cite{bialon2011,Gou2013}). The combined library of 175 structure types calculated for 41 M-B systems under at least 2 pressures contains over 12,000 entries.

We would like to stress that the set of identified \emph{more} stable phases at selected compositions, especially under high pressures, are not necessarily \emph{the} ground states: the aforementioned examples on CaB$_6$~\cite{Kolmogorov2012,shah2013}, WB$_3$/WB$_4$ ~\cite{Zhao2010,Liang2013,Zhang2012}, and FeB$_4$~\cite{Kolmogorov2010,Gou2013} have illustrated that reliable determination of materialÕs ground state requires an enormous number of calculations. In particular, our detailed studies of individual M-B systems \cite{Kolmogorov2010,bialon2011,Kolmogorov2012, Bil2011} showed that the vibrational entropy term in Gibbs energy calculated at elevated temperature can noticeably affect the zero-temperature relative stability and shift the pressure-induced phase transformations by a few GPa. Therefore, we view our findings as a guide for future comprehensive studies of select systems. We also note that since this paper overviews work on metal borides carried out over the last 60+ years, it was necessary to place an emphasis on referencing larger reviews of the materials systems (such as \citet{massalski1990, Rogl1998, rogl1992, Rudy1969}) over studies of individual materials.
 
\section{Structure Determination Methods}

The structures considered within this work were determined in two ways.  The first one was a scan of the Inorganic Crystal Structure Database (ICSD) ~\cite{ICSD1987} for relevant metal-non-metal structure types, critical assessment of available M-B binary phase diagrams\cite{massalski1990} , and an overview of recently published studies on individual M-B systems. Since the presence of (meta)stable B-rich intercalated compounds or M-rich alloys rarely affects the stability of ordered M-B compounds away from the edges of the phase diagram, we did not consider M$_x$B$_{1-x}$ phases with compositions $x<0.08$ or $x>0.85$. 
The second approach was to perform evolutionary ground state searches for selected fixed compositions with Module for \emph{Ab Initio} Structure Evolution (MAISE)~\cite{Kolmogorov2012} software code developed by the authors. The unconstrained optimization was seeded with both random and known structures. Due to the considerable computational cost of density functional theory (DFT) calculations, the evolutionary searches were carried out for a small set of chosen compositions primarily to investigate existing discrepancies between experiment and theory (i.e. W-B).  For a list of compositions see the Supplemental Material~\cite{SupMat}.

Nearly all of the collected 175 structures in our library have been considered for each M-B system. The few exceptions were very large structures, over 40 atoms per unit cell, that were clearly unnatural for certain metal classes. For example, our selected tests showed that complex ground states in alkali and alkaline-earth metal-boron systems had very large positive formation energies for TM-B binaries. The resulting database has been generated in a high-throughput fashion using the MAISE-based framework described in the Supplementary Material~\cite{SupMat}.

Following the determination of structures relevant to M-B systems these structures were studied with DFT.  For this study the thermodynamic stability due to phase separation of each material composition was determined through a construction of the convex hull for each M-B system.  The enthalpy of formation per atom ($H_f^{M_nB_m}$) for each system was determined using:
\begin{equation}
H_f^{M_nB_m}=(H_{tot}-n \mu_{M}-m\mu_{B})/(n+m).
\end{equation}
Here $H_{tot}$ is the total enthalpy of the material composition, $n$ is the number of M atoms, $m$ is the number of B atoms, $ \mu_{M}$ is the chemical potential of the lowest energy bulk metal phase, and $\mu_{B}$ is the chemical potential of $\alpha$-B for 0 GPa and $\gamma$-B for 30 GPa.  For a discussion on our selection of $\alpha$-B for 0 GPa see the authors previous work~\cite{Kolmogorov2010}.  The $H_f^{M_nB_m}$ is then used to construct a convex hull with the stable (metastable) structure defining (being 20 meV/atom above) the corresponding tie line.  These determined stable and metastable structures are then compared to  literature and the ICSD to determine discrepancies between experiment and computation.

\section{Calculation Methods}

All of the energy calculations and structural optimizations, including those within the MAISE software, were performed using the \emph{ab initio} software VASP~\cite{VASP1996}.  The MAISE evolutionary search settings were similar to those described in our previous studies \cite{shah2013,Kolmogorov2010}. The projector augmented-waves~\cite{Kresse1999} formalism was used to treat the core electrons.  All the compounds considered in this study were calculated using the GGA-PBE~\cite{Perdew1996} exchange correlation functional.  The resulting stable and metastable compounds were then calculated using the LDA-PW\cite{perdew1981} functional.  Finally, a few select systems were verified with ultrasoft potentials~\cite{Vanderbilt1990}.  For a majority of the systems the stability or metastability of a compound as determined with the two functionals were the same. The specific differences found between the GGA and LDA results are discussed in more detail in Section \ref{sec:sysdes}.  The same potentials are used regardless of the pressure considered.  A plane wave basis set with an energy cutoff of 500 eV was used with a dense Monkhorst-Pack k-point mesh~\cite{Pack1977,Monkhorst1976} selected specifically for each structure.  Formation energies are expected to be numerically converged to within 1-2 meV/atom. The stress tensor is optimized to the value of the specified pressure  (0 GPa of 30 GPa).  Compounds containing the five magnetic elements (Cr, Mn, Fe, Co, Ni) were calculated both with and without spin polarization.  The known antiferromagnetic elements (Cr and Mn) were also treated with antiferromagnetic initial conditions.  The antiferromagnetic initial conditions were constructed by identifying planes in each of the a$_1$, a$_2$, and a$_3$ directions and alternating the spin either within a plane or between neighboring planes.  For compounds with odd number of magnetic elements super cells of the form 2x2x1 were created to allow for in-plane spin variation.  For the Sc-B system phonon calculations were performed to determine the dynamic stability of the cF52-ScB$_{12}$ \sg{225} phase.  We used the finite displacement methods in the PHON~\cite{Alfe2009} code for these calculations.

\section{Results and Analysis}\label{sec:resanal}

The calculations of the 41 metal boride systems found that around 191 stable phases and that 184 metastable phases exist.  The formula, formation energy, stability (distance to the tie line), and Pearson symbol for both the GGA-PBE and LDA-PW functionals are shown in Tables \ref{tab:stab} (stable compounds) and \ref{tab:meta} (metastable compounds).  The Wyckoff positions of the stable structures (and select metastable structures) are in the Supplemental Materials~\cite{SupMat}.  The metal-boride systems were analyzed for both ambient (0 GPa) and high (30 GPa) pressures.  The changes between 0 GPa and 30 GPa manifest both as same composition phase transitions or as known/new compositions becoming stable/unstable due to phase separation.  Individual analysis of how the M-B systems change with pressure is discussed in Section \ref{sec:sysdes}.  We have also adopted the following naming convention: a \emph{system} denotes a combination of elements, a \emph{compound} specifies a material at a particular composition, a \emph{structure (type)} represents a unit cell and atomic positions for unspecified elements, and a \emph{phase} corresponds to a compound in a particular structure.

\begin{figure}
\centering  
\includegraphics[width=0.45\textwidth]{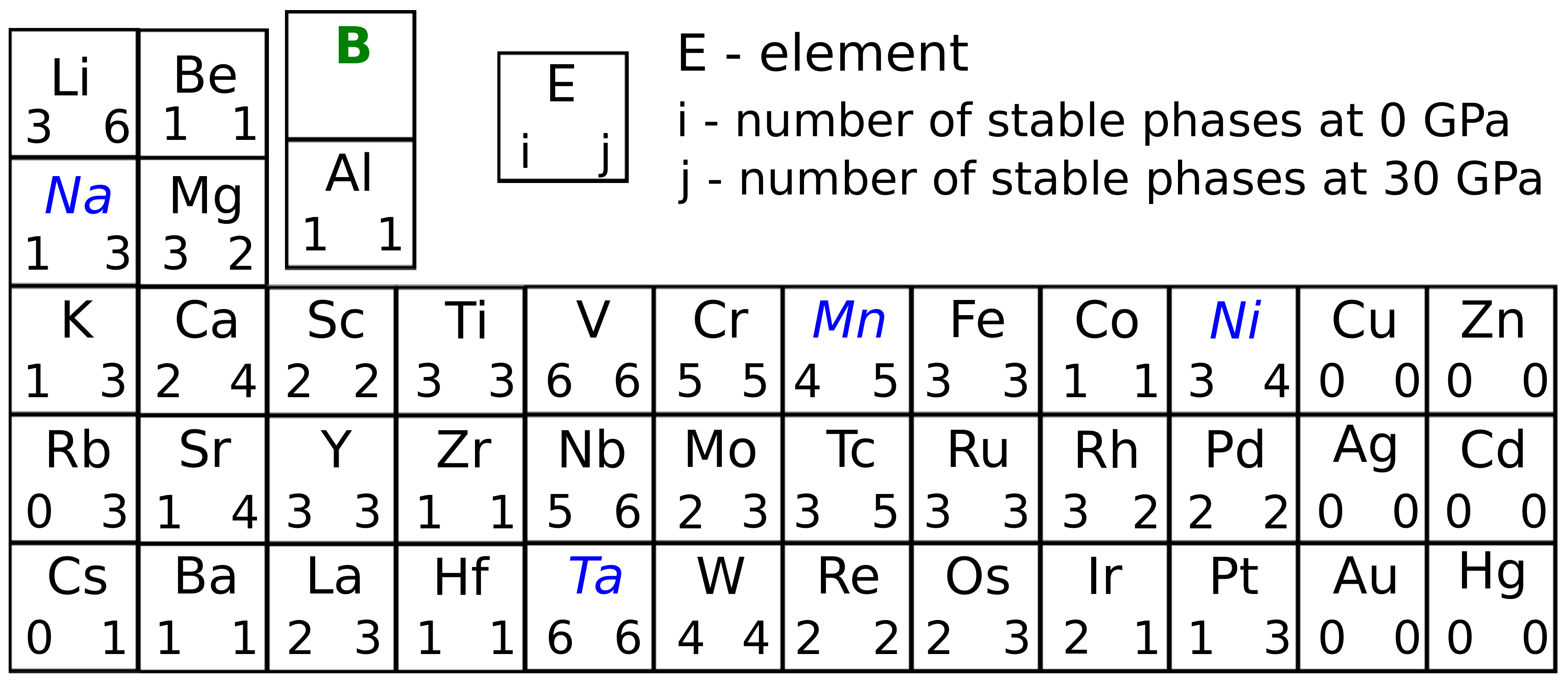}
\caption{The elements considered in this M-B structure study.  Included with each (non-B) element is the number of stable phases at 0 GPa (lower left) and 30 GPa (lower right) for the corresponding M-B system.  B is in green (bold), while blue (italic) represents systems with new phases or structures predicted in this study.}\label{img:PTstab}
\end{figure}

Generation of the large datasets at the two pressures made it possible for us to examine pressure-induced changes in the number or type of stable compounds across the whole set of considered metal boride systems as shown in Fig. (\ref{img:PTstab}). It should be noted though that the selective application of the evolutionary searches disproportionately increased the likelihood of identifying new stable phases under pressure at the studied compositions, e.g, in the Ca-B system. One can distinguish three sets of metal-boron systems showing similar response to applied pressure. For the alkali-metal block, we observed not only an increased number of stable borides under higher pressure but also phase transformation of all stable ambient-pressure compounds with the exception of tP136-Li$_3$B$_{14}$. For the columns IIIB-IVB in the TMs, we did not see any phase changes or change in number of stable phases between the two pressures except for La, which had one more stable compound at 30 GPa. For the columns IB and IIB in the TMs, there were no stable compounds at either pressure. In contrast to the three sets, the alkaline-earth and columns VB-VIIIB in the TMs did not exhibit common distinguishable trends within themselves.  These metal-boron systems were found to have many competing phases which made it difficult to determine the convex hull, let alone count the pressure-induced phase changes. The finite-temperature contributions and DFT systematic errors are likely to have the most impact on the relative stability of close lying phases in these systems and should be investigated in future studies.

\begin{figure}
\centering 
\label{img:hfvn000}\includegraphics[width=0.5\textwidth]{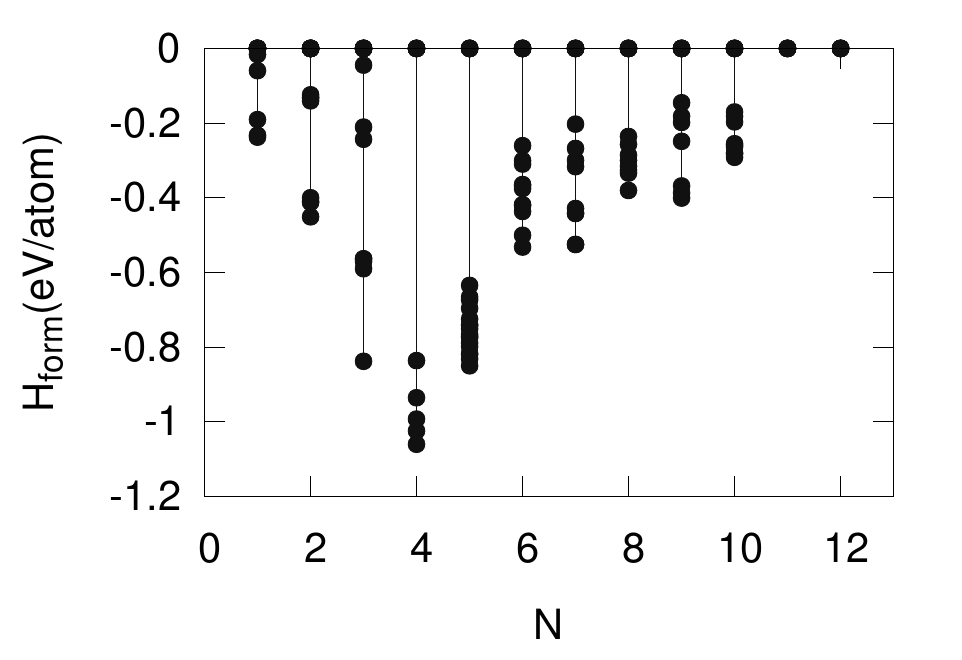}
\caption{Tie lines of each metal boride system at P = 0 GPa versus the electron count (column in the periodic table) of the metal.  A minimum formation enthalpy is observed for $N$ = 4.}\label{img:NvTie}
\end{figure}

In order to rationalize the stability of individual compounds we performed a meta-analysis of the formation energies of all the stable compounds at ambient pressure. Figs.  (\ref{img:alk}-\ref{img:tran5}) show the collection of all individual M-B tie lines and reveal clear general trends in stability as a function of the metalÕs electron count. The formation energies of the compounds in each system starts low for metals with a low valency and decreases in formation energy (becoming more stable) until the metals with a valency of 4 electrons (Ti, Zr, and Hf) is reached after which the formation energy increases until no stable systems exist for columns VB-VIIIB transition metals.  This demonstrates that metal boride systems with metals that are closer to column IVB are more stable with the metals furthest from N=4 being unstable.  Fig. (\ref{img:NvTie}) shows the tie lines of each system combined by column in the periodic table valency and plotted versus formation energy.  The plot shows that the instability of noble metal borides observed previously for selected layered structures~\cite{Oguchi2002, Kolmogorov2006} is a more general trend holding for a much larger set of configurations.  It is worth noting that the Cs-B and Rb-B systems do not contain stable compounds at ambient pressures.  It is expected to be caused by the large atomic size of Cs and Rb, which is supported by the existence of stable structures for Cs and Rb at higher pressures due to compression of boron frameworks around the large atoms.

Interestingly, the Pt-B system has been found to be the only considered M-B binary that destabilizes at elevated pressures. We have checked that the finding is not sensitive to the choice of the exchange-correlation functional (GGA or LDA) or the pseudo-potential (PAW or ultrasoft~\cite{Vanderbilt1990}), i.e., in all these cases the Pt-B compounds under pressure show reduced magnitudes of formation enthalpy. Explanation of this puzzling result will require further study.

\begin{figure}[h]
\centering
\includegraphics[width=0.5\textwidth]{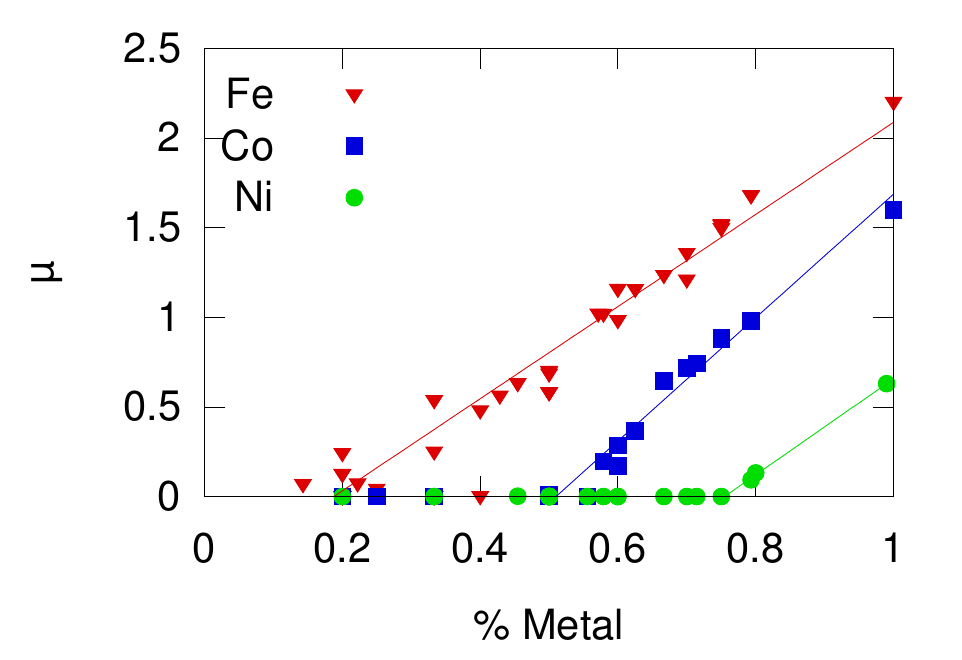}
\caption{A plot of the magnetic moment per atom (boron and metal) versus the metal content of all structures considered with a negative formation energy within 0.15 eV/atom of the tie line for Fe (triangles), Co (squares), and Nickel(circles).  Increasing the cutoff energy above 0.15 eV/atom increases the noise of the data sets without greatly effecting the resulting fits, especially for the 33\% and 50\% compositions, where many structure types were studied.  The fits to these data sets are red for Fe, blue for Co, and green for Ni.}\label{img:FM}
\end{figure}

Another interesting trend observed within this large data set was a nearly linear relationship between the magnetic moment per atom and the metal content above a certain saturation point for the considered ferromagnetic TM-B systems (TM = Fe, Co, Ni) , as seen in Fig. (\ref{img:FM}). Only relevant structures with negative formation enthalpy and within 0.15 eV/atom of the convex hull were taken into account to exclude unrealistic configurations. The decrease in the metal concentration is expected to result in the decrease in the magnetic moment per atom, but the well-defined linear trend is rather surprising given the large variety of structural morphologies in the considered structure set. In fact, we did not observe such a clear correlation when we examined (i) the magnetic moment per \emph{metal} atom as a function of composition; (ii) the magnetic moment per atom as a function of the metal-metal distance; or (iii) the antiferromagnetic moment per atom for any of the Cr, Mn, Fe, Co, Ni - B systems. The saturation point is seen to increase from 20\% metal for Fe to 50\% for Co and 75\% for Ni.  It must be noted for Ni that the limited amount of structures above 80\% metal content reduces the size of the dataset.  However, the results are still consistent across the Fe-Ni set.

The observed correlation can be helpful for the rational design of material properties. For example, the theoretical and experimental results indicate that the recently discovered FeB$_4$ compound is likely a phonon-mediated superconductor. The finding is unusual considering that ironÕs strong itinerant magnetic moment either prevents the superconducting transition altogether or defines it through spin fluctuations~\cite{Mazin2010}. In case of FeB$_4$, the observed $T_c$ of 3 K was much lower than the calculated value of 15-20 K. The original superconductivity calculations indicated that the phonon-mediated $T_c$ would be particularly sensitive to the electron count due to the sharp drop of density of states (DOS) near the Fermi level. If the phonon-mediated scenario is correct electron doping of FeB$_4$ with Co or Ni should increase the $T_c$ considerably by moving the Fermi level into the high DOS peak while not making the compound magnetic, according to  Fig. (\ref{img:FM}).

This extensive study of metal borides has identified several new phases at ambient pressure and many new ones at high pressure. The recently proposed and confirmed phases in our previous studies~\cite{Niu2012,Kolmogorov2012, shah2013,Kolmogorov2010,Kolmogorov2006}  are briefly discussed here but not referred to as ÔnewÕ. Considering that the vast majority of experiments have been carried out under normal conditions, our comparison analysis has been dedicated primarily to ambient-pressure metal borides. Nevertheless, the lattice parameters and Wyckoff positions of all stable and selected metastable calculated materials are included in the Supplemental Material\cite{SupMat} for both 0 GPa and 30 GPA.  4 previously unreported phases at 0 GPa have been calculated to be stable with respect to phase separation. The new ambient pressure materials are summarized in Table \ref{tab:newmat} and classified based upon whether they are a new compound and whether they are new (from MAISE) or previously known structures. Here we discuss these newly predicted structures.

\begin{table}[h]
\caption{The stability of the newly predicted materials by functional and whether the  composition is new and whether the structure was previously known.  The distance to the tie line $dH$ is negative if the new structure is more stable.}\label{tab:newmat}
\begin{tabular}{l   c c  c}
\hline\hline
   Material              &  Compound                &Structure                 & $dH$ (meV/atom)\\
                 &      &  & (GGA/LDA)\\
\hline
 mP20-MnB$_4$ &                Known                            & New                & (-2 / -2)\\
 oS20-Ta$_2$B$_3$&            New                               & Known            & (-2 / -7) \\
 mS28-Ni$_5$B$_2$&            New                              &  Known            &  (-4 / +3)\\
 tP16-NaB$_3$ &              New                                  &  Known            &  (+6 / -3)\\

\hline\hline
\end{tabular}
\end{table}

The exact structure of MnB$_4$ has remained a puzzle ever since the synthesis of the compound in 1970~\cite{Fruchart1960}. It has been argued\cite{Andersson1969,Andersson1970, Burdett1988} that the structure should be a monoclinic distortion of an oI10 \sg{71} configuration originally proposed for CrB$_4$~\cite{Andersson1968}.  Our previous work demonstrated that the oI10-CrB$_4$ \sg{71} and oI10-FeB$_4$ \sg{71} phases are dynamically unstable~\cite{Kolmogorov2010,bialon2011} and later experiments confirmed the derived lower-symmetry oP10 \sg{58} phase~\cite{Gou2013, Knappschneider2013, Knappschneider2011}. Here, the oI10 $\rightarrow$ oP10 distortion was found to lower the energy of MnB$_4$ as well (by 10 meV/atom). Additional evolutionary searches have produced even more stable configurations related to oI10, e.g., an antiferromagnetic oF80-MnB$_4$ \sg{43} phase 6 meV/atom below oP10. A new non-magnetic mP20 \sg{14} structure, 8 meV/atom below oP10, is proposed to be the ground state structure for MnB$_4$. The monoclinic structure of MnB$_4$ was recently independently described by Bykova and co-authors based on unpublished single crystal X-ray diffraction data~\cite{MnStruc}.  Table I in the Supplemental Material~\cite{SupMat} contains the cell parameters and Wyckoff position of this compound.

Three additional compounds are proposed here in the Ta-B, Ni-B, and Na-B systems (see Supplemental Material for cell sizes and Wyckoff positions~\cite{SupMat}). The oS20-Ta$_2$B$_3$ \sg{63} phase of the V$_2$B$_3$ prototype is seen just below the tie line (by 2 meV/atom) defined by hP12-TaB$_2$ \sg{194}  and oI14-Ta$_3$B$_4$ \sg{71}~\cite{Kiessling1949, Okada1993}.  We noted an unusual sensitivity of the relative energy of competing TaB$_2$ polymorphs to the choice of the DFT approximation: hP3 \sg{191}  is below hP12 by 15 meV/atom in the GGA while hP12 is below hP3 by 12 meV/atom in the LDA (and literature ~\cite{Norton1949, Okada1993}). However, in both cases oS20-Ta$_2$B$_3$ \sg{63} appears as marginally stable. The proposed Ni-B compound is mS28-Ni$_5$B$_2$ \sg{15}, 4 meV/atom below the tie line defined by tI12-Ni$_2$B \sg{140}~\cite{Bjurstroem1933} and oP16-Ni$_3$B \sg{62}~\cite{Rundqvist1967} in the GGA and 3 meV/atom above the tie line in the LDA. Considering the (near) stability shown with the two functionals, the compound is a viable candidate to exist.  The identified new tP16-NaB$_3$ \sg{127}  phase (of the tP16-Li$_2$B$_6$ prototype) is calculated to be stable with the LDA functional at 3 meV/atom below the tie line defined by oS46-Na$_3$B$_20$ \sg{65}  structure and bulk Na.  However, the GGA-PBE functional calculates NaB$_3$ to be 6 meV/atom above this tie line.

Along with these newly proposed compounds we were able to confirm another compound that was recently proposed in literature.  \citet{Liang2013} used an evolutionary search for the WB$_3$ composition and found an hR24 \sg{166} structure to be more stable than the hP16 \sg{194}  structure previously considered~\cite{Zhao2010, Liang2012, Liang2011,Zhang2012}.  Our calculations based on both high-throughput and evolutionary searches confirm that the hR24 structure is 8 meV/atom more stable than the hP16 structure.  In fact, this structure is one of the three metastable ones that was predicted by \citet{Zhang2010} in 2010 during their detailed evolutionary search of the Mo-B system.  The Mo-B and W-B systems are discussed in more detail in Section \ref{sec:sysdes}.

This study also provides evidence against both theoretically and experimentally predicted/reported structures. Both hP3-AuB$_2$ \sg{191}  and hP3-AgB$_2$ \sg{191}  phases have been reported~\cite{Obrowski1961}. However, the stability of these structures has been questioned either due to the difficulty of synthesis and the instability over extended periods~\cite{Islam2007,tobin2008} or due to positive formation energies~\cite{Kolmogorov2006}.  Our calculations demonstrate that none of the noble metal - boron systems contain any stable phases.  Another reported phase we found unstable was mP16-NiB$_3$ \sg{14} proposed by \citet{Caputo2010}.  The mP16 structure, generated from the parameters in Ref. \cite{Caputo2010}, relaxed by 283 meV/atom in our VASP calculations, but its final formation energy remained positive at 0.132 meV/atom. Our evolutionary search at the NiB$_3$ composition identified a considerably more stable mS32 \sg{12} structure, 179 meV/atom below mP16, but still 77 meV/atom above the tie line.  Finally, tP20-RbB 4 \sg{127} (of the UB 4 prototype) has been predicted as a previously unobserved stable phase \cite{Chepulskii2009}.  However, our calculations show this structure to be unstable with a positive formation energy of 0.253 eV/atom.

The overall correspondence between the calculated and observed sets of stable compounds is rather difficult to quantify. If we compare the number of experimentally seen phases (excluding intercalated compounds, those that are too large to be simulated with DFT, and high temperature or high pressure polymorphs) with the number of phases that are calculated with either LDA or GGA to be stable we see that 75.5\% of the experimental phases are stable in the DFT.  The number increases to 77.2\% if we also include, as in Ref. \cite{Curtarolo2005a} , the agreement between observed and calculated immiscibility. Each of the 8 immiscible systems correctly found to be immiscible was counted, conservatively, as one hit (Au and Ag are considered as immiscible systems due to the lack of a reported stable structure in the literature).  However, this number might not accurately reflect the discrepancies between our DFT calculations and experiment because there are 4 newly predicted phases here.  Therefore, if we instead track the number of discrepancies between theory and experiment and divide it by the total number of compounds (those reported experimentally and those newly predicted here) the agreement rate is 76.3\%.  It is important to note that both experiment and computation contribute to the error in this agreement rate. Therefore, to gain a better understanding of the computational error, we removed the discrepancies that deviated from the tie line by more than 50 meV/atom. Since this cutoff is higher than the expected numerical/systematic errors and the missing finite temperature contributions, these disagreements are likely due to formation of truly metastable phases or errors in the interpretation of the experimental data. This adjustment increases the agreement rate to 83.6\%.  When the experiment is similarly compared to just GGA (LDA) the percentage is 79.6\% (81.6\%).

The estimated agreement rate of 83.6\% (75.2\%) obtained for the metal borides excluding (including) non-computational discrepancies is lower than the corresponding value of 97.3\% (92.4\%) reported by \citet{Curtarolo2005a} for 80 metal alloys. The lower agreement percentage seen in this metal boride study is not unexpected considering that (i) the synthesis/characterization of metal borides is complicated by their structural complexity and high melting temperatures and (ii) the DFT errors for materials relative stabilities are more likely to cancel out for metal alloy structures, which have more closely matched charge densities.  


\section{Conclusions}

In summary, the resulting dataset for ambient-pressure $s$-$p$ and TM boride systems:

\begin{itemize}

\item demonstrates that the formation energy of the stable compounds of a system is lowest for metals with a valence of 4 with the stability of the systems decreasing on either side of this valence until the noble metals are completely unstable;

\item demonstrates a linear relationship between the magnetic moment per atom and the metal content for the ferromagnetic Fe, Co, and Ni-based compounds with negative formation energy within 0.15 eV/atom of the system's tie line; the linear fits give 20\%, 50\%, and 75\% as minimal values of the Fe, Co, and Ni content for the relevant compounds to be magnetic;

\item includes one proposed new mP20-structure \sg{14} (as the true ground state of MnB$_4$) and three new candidate compunds (oP10-Ta$_2$B$_3$ \sg{63}, mS14-Ni$_5$B$_2$\sg{15}, and tP16-NaB$_3$  \sg{127});

\item serves as a benchmark for the DFT calculations of crystal structure stability of 83.6\% agreement between calculated and expected stable metal borides.

\item singles out Pt-B as the only considered system with pressure-induced \emph{destabilization} of binary compounds.

\end{itemize}

Finally, this study illustrates the need to go beyond the perviously known structures to correctly characterize each of the metal boride systems and to search for new materials.  The study of column VB-VIIIB  TM borides at ambient pressure is a strong example of this.  However, it is within the high pressure regime that opportunities for the discovery and prediction of new materials are particularly likely.  

\section{System Descriptions}\label{sec:sysdes}

\subsection{$s$-$p$ metals}

\begin{figure*}[!htp]
\centering 
\includegraphics[width=0.885\textwidth]{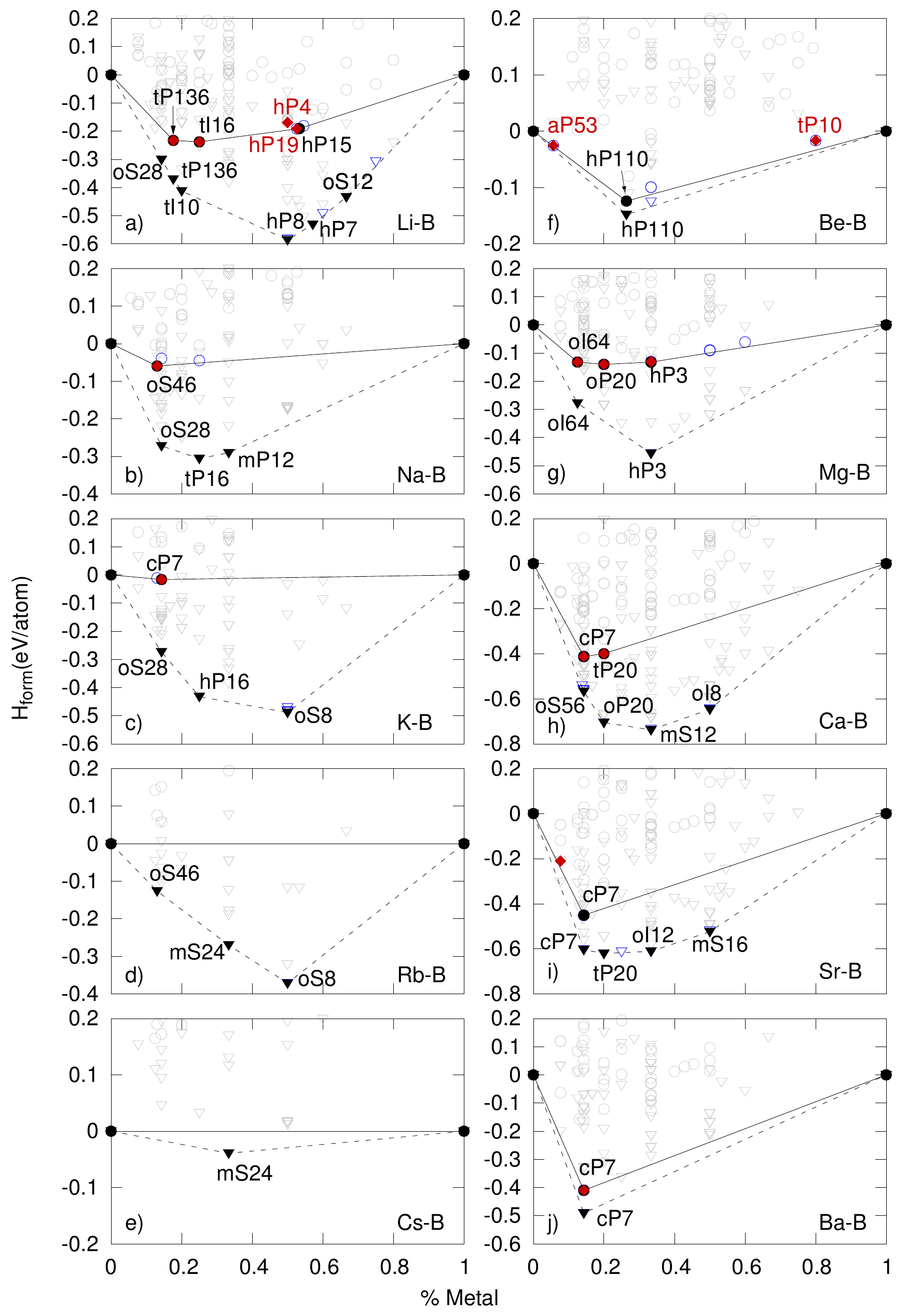}
\caption{The tie lines and convex hulls of the alkali (Li through Cs) and alkali-earth (Be through Ba) metals.  Circles are 0 GPa calculations, triangles are 30 GPa calculations, and diamonds are 0 GPa reported experiments.  Black shapes are stable, gray shapes are unstable, and blue shapes are marginally-stable systems.  Black labels are stable Pearson symbols from this work, while red labels are the Pearson symbols from experiment.  If the label corresponds to both this work and literature the label is black.}\label{img:alk}
\end{figure*}

The convex hulls for the alkali, and alkaline-earth metals are in Fig. (\ref{img:alk}), while the convex hull for Al is in Fig. (\ref{img:tieAl}).

\subsubsection{Li-B}

\citet{Vojteer2009} has stated that three phases of Li-B have been well quantified with many other phases compositions being either undetermined or undeterminable. The three quantified structures correspond to the Li$_2$B$_6$, Li$_3$B$_{14}$, and Li$B_{0.88}$ compositions. The boron framework comprised of B$_6$ octahedra in $M_n$B$_6$ is typically stabilized by a single large metal atom per cP7 unit cell ($n=1$). Two small Li$^+$ ions fit into each cavity causing the $B_6^{2-}$ units to rotate and break the cubic symmetry. The resulting tetragonal structure \sg{127} has been observed to host Li in the $4h$ and $4f$ Wyckoff sites with 0.80 and 0.20 occupancies, respectively~\cite{Schnering1999}. As in the study of \citet{Chepulskii2009}, we simulated the structure with the majority $4h$ sites fully occupied and confirmed the stability of the tI16-Li$_2$B$_6$ \sg{127} phase. The tI160-Li$_3$B$_{14}$ \sg{122} phase based on B$_8$ and B$_{10}$ units~\cite{Mair1988} has multiple Li sites with fractional occupancies of 0.50 or 0.40. We were able to retain the 3:14 composition in a tetragonal tP136 cell \sg{81} by populating 50\% of metal sites and picking an atomic arrangement with the most natural Li-Li distances. The simulated phase was found to be soundly stable without the inclusion of the configurational entropy contribution. Determination of the morphology and the exact composition of the nearly stoichiometric LiB$_x$ compound required a set of extensive studies discussed and performed by \citet{Kolmogorov2006}. The simplest proposed representations of the linear boron-chain structure were $\alpha$-LiB~\cite{Liu2000,Worle2000} and $\beta$-LiB~\cite{Rosner2003}. However, the structural models are not suitable for interpretation of powder XRD data as the fitting leads to unphysically short B-B distances. An insightful solution proposed by W\"orle and Nesper in 2000~\cite{Worle2000} was that the off-stoichiometry is a result of incommensurability of the B and Li sub-lattices. A computational study of LiB$_x$ by  \citet{Kolmogorov2006} illustrated that maximum stability occurs at compositions close to $x\approx0.90$ in excellent agreement with experiment. An established linear dependence between $x$ and the easy-to-measure $c_{Li-Li}$ provided a simple recipe for monitoring the compound's composition and suggested that the $x=0.8-1.0$ range of stability is greatly overestimated.  An unusual feature of the Li-B phase diagram near the 1:1 composition is that LiB$_x$ has a small but finite range of stability at $T=0$ K and may be changed post-synthesis~\cite{Kolmogorov2007}. The commensurate hP15-Li$_{8}$B$_7$ \sg{187} representation of LiB$_x$ defines the convex hull in this work.

At $P=30$ GPa our DFT calculations show that 6 structures are stable. In agreement with our previous study~\cite{Kolmogorov2006}, the region of LiB$_x$ stability shifts further away from 1:1 with pressure and hP7-Li$_4$B$_3$ \sg{187} takes the place of Li$_{8}$B$_7$. The tP136-Li$_3$B$_{14}$ compound retains the same structure, while the hP15-LiB$_3$ phase becomes unstable and is replaced by the tI10-LiB$_4$ \sg{139} phase. Among the considered structures on the B- and Li-rich sides, new oS28-LiB$_6$ \sg{65} and oS12-Li$_2$B \sg{63} compounds become stable. The 1:1 ground state is a previously proposed 'metal sandwich' hP8-LiB \sg{194} phase. This stoichiometric layered phase was predicted to be a close superconducting analog to MgB$_2$~\cite{Kolmogorov2006b}. The first attempts to synthesize it were unsuccessful, but recent computational studies~\cite{Peng2012, Hermann2012, Hermann2013} have reproduced the stability of the proposed 'metal sandwich' phases at this pressure. The investigations of the Li-B system under higher pressures~\cite{Peng2012, Hermann2012, Hermann2013} have revealed a number of interesting ground states across the full composition range.

\subsubsection{Na-B}

The Na-B system has had several phases suggested including oS46-Na$_3$B$_{20}$  \sg{65}~\cite{Albert1999b}, oS28-NaB$_6$~\cite{massalski1990}, NaB$_{15}$~\cite{Naslain1970}, NaB$_{16}$~\cite{massalski1990}, and mS64-Na$_2$B$_{29}$  \sg{8}~\cite{albert2000}.  However, \citet{albert2000} has shown that Na$_2$B$_{29}$ is the correct composition for the NaB$_{15}$ and NaB$_{16}$ compounds, which are known to be stable at non-zero temperatures.  Moreover, they are intercalated structures so are not considered in our calculations.  The NaB$_6$ compound is now known to only be stable with the inclusion of C to form NaB$_5$C~\cite{albert1999, Orlovskaya2011} or as the oS46-Na$_3$B$_{20}$  \sg{65} phase~\cite{Albert1999b, Albert2000b}.  Therefore, the stable Na$_3$B$_{20}$ compound seen here agrees with the known experimental ground state. However,  our LDA calculations indicates a possibly stable phase, tP16-NaB$_3$ \sg{127}, which is of the Li$_2$B$_6$ prototype.  

The ambient-pressure oP23-Na$_3$B$_{20}$ phase becomes unstable at higher pressure giving way to the more common nearby MB$_6$ composition, implying that both C inclusion and increased pressure can stabilize the NaB$_6$ phase.  Also, at 30 GPa two other phases are calculated to be stable, tP16-NaB$_3$  \sg{127} and mP12-NaB$_2$ \sg{10}.

\subsubsection{K-B}

Experimentally, the K-B system contains a single phase, cP7-KB$_6$ \sg{221}~ \cite{Naslain1966}. This agrees with our calculations, although the phase's formation energy is found to be barely negative  (see Fig. \ref{img:alk}).  Application of high pressure  is expected to induce phase transformation, as one of the metastable structures found with MAISE for CaB$_6$,  oS28-\sg{65}, stabilizes over cP7.  Further, two new phases form at 30 GPa, namely the hP16-KB$_3$ \sg{194} and oS8-KB \sg{63}.  The oS8 structure has a common morphology for this M-B composition composed of  linear B chains.

\subsubsection{Rb-B}

The Rb-B system contains no intermediate phases as predicted in our calculations at ambient pressures.  However, at 30GPa oS46-Rb$_3$B$_{20}$ \sg{65}, mS24-RbB$_2$ \sg{12}, and oS8-RbB \sg{194} are all seen to be stable.

\subsubsection{Cs-B}

The Cs-B system contains no intermediate compounds, which matches our results.  Within our calculations, at P=30 GPa,  mS24-CsB$_2$ \sg{12} is the only phase observed (for both pressures) to have a negative formation energy after running an evolutionary search.

\subsubsection{Be-B}

The Be-B system is a complex system with many metastable and uncertain phases~\cite{massalski1990}, especially within the high boron content compositions. \citet{Walsh2009} identified the following set of Be-B phases: cF12-Be$_2$B \sg{225}, hP117-BeB$_2$ \sg{191}, tP196-BeB$_6$ \sg{76}, and possibly tP10-BeB$_4$ \sg{129}.   \citet{Hermann2012} used ab initio techniques to explain the uncertainty in the literature about the boron-rich Be-B phases in the 20-33\% composition range. They found the BeB$_{\sim2.75}$ compound, represented as hP110-Be$_{29}$B$_{81}$ \sg{187}, to be the most stable within this region with other slight variations in stoichiometry possible due to partial occupancy of Be sites.  In agreement with the follow-up paper of \citet{Hermann2012}, we observe hP110-Be$_{29}$B$_{81}$ to be the only stable Be-B compound with tP10-Be$_4$B \sg{129} and aP53-Be$_3$B$_{50}$ \sg{1} metastable by 17 meV/atom and 1 meV/atom, respectively. We also had constructed, using chemical reasoning, a cF12-BeB$_2$ \sg{216} phase with the diamond B$^{1-}$ network stabilized by the insertion of the small Be$^{2+}$ cations, which has been recently proposed by \citet{Hermann2012} as the lowest-energy BeB$_2$ structure at 0 GPa and stable at very high pressures~\cite{Hermann2013}. However, our evolutionary search at 0 GPa has uncovered a considerably more stable  oS12 \sg{63} structure, 85 meV/atom below cF12 and only 13 meV/atom above the tie line. The new structure (see the Supplemental Materials~\cite{SupMat}) is comprised of hexagonal buckled B sheets with Be sitting above the middle of a B hexagon in one layer and on top of a B-B bond in another. The high-temperature cF12-Be$_2$B \sg{225} structure seen in the literature is unstable by 44 meV/atom in our calculations. At 30 GPa we calculate that the hP110-Be$_{29}$B$_{81}$ structure is the only stable structure in agreement with \citet{Hermann2013} who predicted the first changes in the Be-B convex hull occurring between 20 and 80 GPa.

\subsubsection{Mg-B}

The known Mg-B ground states are the hP3-MgB$_2$ \sg{191}~\cite{Jones1954}, oP20-MgB$_4$ \sg{62}~\cite{Guette1972}, oI64-MgB$_7$ \sg{74}~\cite{Pediaditakis2010} phases, perfectly matching the calculated set.  The theoretical metal sandwich structures~\cite{Kolmogorov2006, Kolmogorov2006b} were originally identified for the Mg-B system.  Both MgB and Mg$_3$B$_2$ composition with additional layers of Mg are metastable at 0 GPa, and neither is stabilized at higher pressures.  Therefore, the only seen change at 30 GPa is that oP20-MgB$_4$ becomes unstable.

\subsubsection{Ca-B}

The ambient and high-pressure Ca-B phases have been systematically explored in recent studies~\cite{shah2013,Kolmogorov2012}. Under normal conditions, cP7-CaB$_6$ \sg{221} and tP20-CaB$_4$ \sg{127} have been observed experimentally~\cite{Pauling1934, liu2010, Johnson1961}. Stability of CaB$_4$ has been the subject of debate~\cite{Schmitt2006}, but the recent~\cite{shah2013,Chepulskii2009} and the present calculations indicate that the compound is thermodynamically stable at 1 bar.

Application of gigapascal pressure leads to the appearance of a number of new ground states. At the 1:6 composition alone, the cP7 structure becomes dynamically unstable and gives way to unexpectedly complex oS56 \sg{63} and tI56 \sg{139} structures~\cite{Kolmogorov2012}. The latter was supported by powder XRD data and is one of the largest structures found without \emph{any} structural input from experiment~\cite{Kolmogorov2012}. The most recent DFT analysis showed that addition of B stabilizes the parent tI56-CaB$_6$ phase \sg{139} further~\cite{shah2013}. One of the possible derived phases, tP57-CaB$_{6.125}$ \sg{123}, could be the ground state for pressures above 32 GPa~\cite{shah2013}. At the 1:4 composition, the metallic ThB$_4$-type structure destabilizes with respect to the semiconductive MgB$_4$-type structure. New stable superconducting compounds appear at the 1:2 and 1:1 compositions~\cite{shah2013}. Our resulting convex hull at 30 GPa is defined by oS56-CaB$_6$\sg{63}, oP20-CaB$_4$ \sg{62}, mS12-CaB$_2$ \sg{71}, and oI8-CaB \sg{74}.

\subsubsection{Sr-B}

The Sr-B system is known to contain the cP7-SrB$_6$ \sg{221} compound~\cite{Ott1997}, as seen in our calculations in Fig. (\ref{img:alk}i).  The cP7-SrB$_6$ \sg{221},  tP20-SrB$_4$  \sg{127},  oI12-SrB$_2$  \sg{71}, and mS16-SrB  \sg{15} structures are all calculated to be stable structures at 30 GPa. Established correlations between the metal ion size and the structure stability for boron-rich compounds~\cite{shah2013} indicate that the large-size Sr and Ba ions will keep the known cP7 structure stable up to at least 40 GPa with respect to considered polymorphs.

\subsubsection{Ba-B}

The Ba-B system has been experimentally reported to have cP7-BaB$_6$ \sg{221} as its only stable compound~\cite{Schmitt2001}.  This matches the cP7-BaB$_6$ \sg{221} phase from our database at ambient pressures and 30GPa.   As discussed for Sr-B, the cP7 structure is particularly stable due to the large size of the Ba ion~\cite{shah2013}.

\subsubsection{Al-B}

\begin{figure}[h]
\centering
\includegraphics[width=0.375\textwidth]{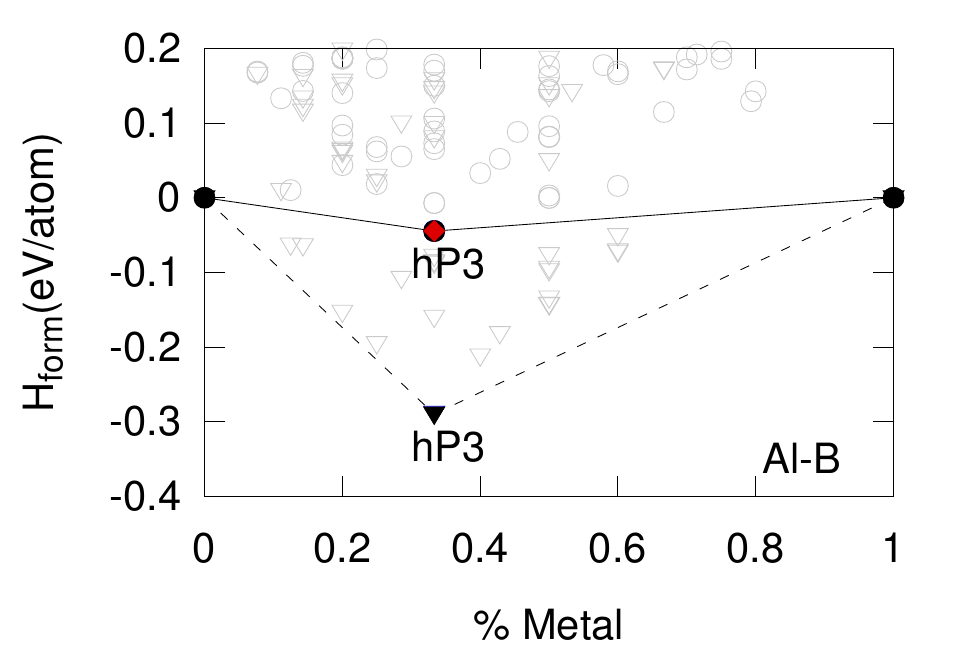}
\caption{The convex hull of the Al-B system.  Circles are 0 GPa calculations, triangles are 30 GPa calculations, and diamonds are 0 GPa reported experiments.  Black shapes are stable, and gray shapes are unstable.  Black labels are stable Pearson symbols from this work.  If the label corresponds to both this work and literature the label is black.}\label{img:tieAl}
\end{figure}

Al is the only considered non-alkali/alkaline-earth $s$-$p$ metal. Our Al-B  convex hull is in Fig. (\ref{img:tieAl}), while the most recent phase diagram is from \citet{Duschanek1994}.  Observed Al-B phases include hP3-AlB$_2$  \sg{191}~\cite{Felten1956}, oC88-AlB$_{10}$  \sg{60}~\cite{massalski1990}, tP216-AlB$_{12}$  \sg{94}~\cite{Duschanek1994} ($\alpha$),  AlB$_{12}$ \sg{74}~\cite{massalski1990} ($\beta$), and oP384-AlB$_{12}$  \sg{19}~\cite{massalski1990}\cite{Duschanek1994} ($\gamma$).  However, both AlB$_{10}$, and $\beta$ AlB$_{12}$ are high temperature phases\cite{massalski1990}.  Our calculations do not include the AlB$_{10+}$ structures due to our chosen limits on intercalation.  Therefore, we can only comment upon the stability of hP3 structure, which appears stable at both ambient and high pressures.

\subsection{Row 3 transition metals}

\begin{figure*}
\centering  

\includegraphics[width=0.885\textwidth]{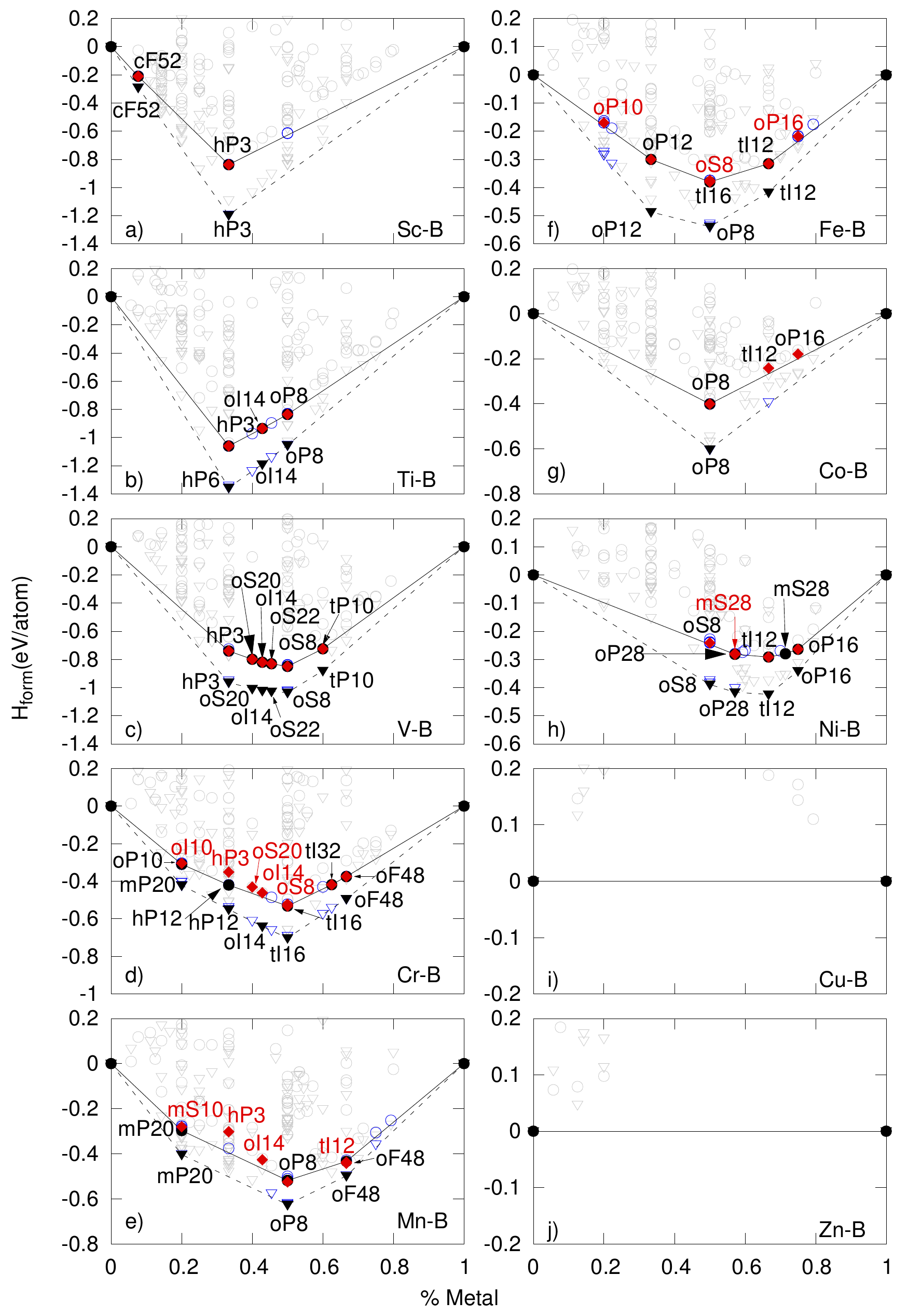}

\caption{The tie lines and convex hulls of the 3$d$ TM (Sc through Zn).  Circles are 0 GPa calculations, triangles are 30 GPa calculations, and diamonds are 0 GPa reported experiments.  Black shapes are stable, gray shapes are unstable, and blue shapes are marginally-stable systems.  Black labels are stable Pearson symbols from this work, while red labels are the Pearson symbols from experiment.  If the label corresponds to both this work and literature the label is black.}\label{img:tran3}
\end{figure*}

The convex hulls of the 3$d$ TMs are in Fig. (\ref{img:tran3}).

\subsubsection{Sc-B}


The Sc-B system is known to have hP3-ScB$_2$ \sg{191} as a stable phase~\cite{Zhuravlev1958}, which agrees with our database. ScB$_{15}$  \sg{76} and ScB$_{19-19.5}$ \sg{92}  have been observed experimentally~\cite{Pediaditakis2011,Tanaka1998}, but are not simulated here due to the large sizes of the intercalated structures. At the ScB$_{12}$ composition, two related\cite{massalski1990} intercalated structures have been reported: cF52-ScB$_{12}$ \sg{225}~\cite{Przybylska1963} and tI26-ScB$_{12}$ \sg{139}~\cite{Matkovich1965}. According to our calculations at 0 GPa, cF52 is thermodynamically stable and its tI26 derivative relaxes back to the cubic configuration. At 30 GPa, cF52 is found to have multiple imaginary phonon modes which makes the identification of nearby dynamically stable derivatives a challenging problem as discussed in Ref. \cite{shah2013}. The proposed 'metal sandwich' hP8 structure~\cite{Kolmogorov2006b,Kolmogorov2006} is only 16 meV/atom above the tie line at 0 GPa but becomes less stable at higher pressures. The 1:12 and 1:2 phases define the convex hull at 30 GPa as well.

\subsubsection{Ti-B}

The Ti-B system is a well understood system with three compounds reported.  These oP8-TiB \sg{62}~\cite{Decker1954}, oI14-Ti$_3$B$_4$ \sg{71}~\cite{massalski1990}, and hP3-TiB$_2$ \sg{191}~\cite{Norton1949} phases all match well with those calculated in this study and shown in the phase diagram of \citet{Nakama2009}. The phases remain stable at 30 GPa.

\subsubsection{V-B}

The V-B system is also a well understood system with a large number of reported stable compounds. The hP3-VB$_2$ \sg{191}, oS20-V$_2$B$_3$ \sg{63}, oI14-V$_3$B$_4$ \sg{71}, oS22-V$_5$B$_6$ \sg{65}, oS8-VB \sg{63}, and  tP10-V$_3$B$_2$  \sg{127} phases have been seen experimentally~\cite{Norton1949, Yu1995,Spear1969, Nowotny1958, Schob1965, Effenberg2009, Riabov1999, Nunes2004} and all agree with those obtained in our database. The set illustrates nicely the evolution of the boron bonding morphology from the flat 2D sheets in VB$_2$, to mixtures of strips and chains for the intermediate V-B compositions, to purely 1D chains in VB, and finally to unlinked sites in V$_3$B$_2$. The perfect agreement between theory and experiment can be attributed to the cancellation of errors in the calculated relative stabilities for the series of closely related structures. The set of stable V-B phases is unchanged at 30 GPa. 

\subsubsection{Cr-B}

Our review of the Cr-B system reveals several discrepancies between theory and experiment and highlights the need for further analysis of some reported Cr-B compositions. The theory-guided revision of the CrB$_4$ structural model illustrates the value of expanding crystal structure searches beyond the known prototype databases.  \citet{Andersson1968} synthesized CrB$_4$ over 40 years ago, and based on powder diffraction data, solved its structure as oI10, which was used until very recently\cite{ Knappschneider2011}. \citet{Kolmogorov2010} and \citet{bialon2011}Õs calculations showed oI10 \sg{71} to be dynamically unstable and predicted a derived oP10 \sg{58} structure with a considerably distorted boron network to be the true ground state. The brand-new oP10 structure has indeed been confirmed for CrB$_4$ by recent experiments\cite{Niu2012, Knappschneider2013} (see the Fe-B section for more related information). The tI32-Cr$_5$B$_3$ phase reported in literature~\cite{Portnoi1969,Bertaut1953,Guy1976} is stable in our database. However, CrB$_2$ has been reported to adopt the hP3 \sg{191} configuration (the AlB$_2$ prototype with flat boron layers)~\cite{Post1954,Guy1976}, while our calculations indicate that the structure is unstable by 70 meV/atom with respect to hP12 \sg{194} (the WB$_2$ prototype with a mixture of flat and buckled boron layers). The majority of literature on CrB reports an oS8 \sg{63} structure~\cite{Frueh1951, Pradelli1976, Okada1987, massalski1990,rogl1992} with the exception of \citet{Papesch1973} who observed a tI16 \sg{141} structure. According to our calculations, tI16 is the ground state with oS8 being metastable by 10 meV/atom. The oI14-Cr$_3$B$_4$ \sg{71} composition of the Ta$_3$B$_4$ prototype has been seen in literature~\cite{Elfstroem1961,massalski1990}, but is found to be 22 meV/atom above the tie line in our calculations matching the experimental expectation of it being a high temperature phase~\cite{rogl1992}.  However, this structure becomes stable at 30 GPa within our database.  Originally, the Cr$_2$B structure was reported to be tI12 \sg{140}~\cite{Bertaut1953}, but has since been described to be oF48-Cr$_2$B \sg{70}~\cite{Guy1976, massalski1990, Lugscheider1974}.  The latter structure matches both the GGA and LDA calculations of this study. Cr$_2$B$_3$ was reported by ~\citet{Okada1987} as oS20 \sg{63}, which is unstable (by 34 meV/atom) in our calculations.  It is worth noting that all of the stable structures in the Cr-B system seen within our DFT calculations were nonmagnetic.  At 30 GPa, our calculations show that mP20-CrB$_4$ \sg{58}, hP12-CrB$_2$ \sg{194}, oI14-Cr$_3$B$_4$  \sg{71}, tI16-CrB  \sg{141}, and oF48-Cr$_2$B  \sg{70} are the stable phases.

\subsubsection{Mn-B}

The Mn-B system, just as Cr-B, contains a series of discrepancies between theory and experiment that suggest the need for further investigation.  The most recent reinvestigation of the full Mn-B system was from \citet{Smid1989}.  The oF40-Mn$_4$B \sg{70} phase reported in Ref. ~\cite{Kiessling1950} has since been described as a vacancy structure of orthorhombic Mn$_2$B \sg{70}~\cite{massalski1990,Tergenius1981, DictionaryInorganic}. An additional high-temperature tI12 \sg{140} structure has been reported in the literature for the Mn$_2$B composition \cite{Havinga1972}. Our calculations support these experimental findings placing the tI12-Mn$_2$B \sg{140} polymorph 5 meV/atom above the oF48-Mn$_2$B \sg{70} ground state at $T$ = 0 K. The calculated stability of oP8-MnB  \sg{62} is also consistent with experiment~\cite{Kiessling1950}. However, the experimentally observed oI14-Mn$_3$B$_4$ \sg{71}~\cite{Kiessling1950} phase is 45 meV/atom above the tie line in our DFT calculations.  The experimental hP3-MnB$_2$ \sg{191} (AlB$_2$ prototype)~\cite{massalski1990, Aronsson1960c} phase is also determined to be unstable by nearly 100 meV/atom. The hP6 \sg{194} (ReB$_2$ prototype) structure is found to be far more energetically favorable than hP3, in agreement with previous DFT calculations ~\cite{Aydin2009}, but it is not clear whether or not hP6 is really stable being 20 meV/atom above and 10 meV/atom below the tie line in the GGA and the LDA, respectively. Our evolutionary search at the MnB$_4$ composition produced a new mP20 \sg{14} structure which we propose to be the true ground state. This finding appears to agree with the results of an independent experimental study by Bykova and co-authors based on unpublished single crystal X-ray diffraction data~\cite{MnStruc}.   Sections \ref{sec:resanal}, Cr-B, Fe-B, and Ref. \cite{Kolmogorov2010} give more information on the relationship between the other related mS10 \sg{12} ~\cite{Andersson1970}, oI10 \sg{71} \cite{Andersson1968}, and oP10 \sg{58} ~\cite{Kolmogorov2010, bialon2011} structures proposed for the MnB$_4$, CrB$_4$, and FeB$_4$/CrB$_4$ compounds, respectively. At 30 GPa mP20-MnB$_4$ \sg{14}, oP8-MnB \sg{62}, and oF96-Mn$_2$B \sg{70} are stable in our study.

\subsubsection{Fe-B}

The Fe-B system has been overviewed and explored in the author's previous studies~\cite{Kolmogorov2010, bialon2011}. Two stable compounds, oP8-FeB \sg{53} and tI12-Fe$_2$B \sg{140}, appear in the experimental phase diagram~\cite{Okamoto2004}, but synthesis of metastable FeB$_{49}$~\cite{Balani2006}, a solid solution of 5\% Fe in B (FeB$_{19}$)~\cite{Callmer1976},  and  Fe-C analogs, cF116-Fe$_{23}$B$_6$ \sg{225} and tI32-Fe$_3$B \sg{82}, have also been reported~\cite{Lanier1994,Watanabe1983,Khan1982}. In good agreement with experiment, our DFT calculations indicate stability of tI12-Fe$_2$B \sg{140} and metastability of cF116-Fe$_{23}$B$_6$ \sg{225} and oP16-Fe$_3$B \sg{62} by 19 meV/atom and 18 meV/atom, respectively.  Three phases of Fe$_3$B,  oP16-Fe$_3$B \sg{62} prototype Fe$_3$C), tI32-Fe$_3$B \sg{82} (prototype Ni$_3$P), and tP32-Fe$_3$B \sg{86} (prototype Ti$_3$P), have been reported with the Fe$_3$C prototype determined to be a metastable phase, while the Ni$_3$P and Fe$_3$P prototypes are high temperature phases~\cite{Khan1982}.  This agrees with our results of  24 meV above the tie line for tI32, 104 meV for tP32, and 18 meV  for oP16.   As discussed by \citet{Kolmogorov2010}, two experimentally observed FeB polymorphs, oP8 \sg{62} and oS8 \sg{63} are slightly above (by at least 5 meV/atom) a tI16 structure in the GGA, but are favored (by at least 10 meV/atom) in the LDA. The evolutionary ground state search of \citet{Kolmogorov2010} unexpectedly produced two viable ground states at 1:2 and 1:4 compositions. The predicted oP12 structure for FeB$_2$ \sg{62} comprised of B chains rather than B layers was found to be over 30 meV/atom below the $\alpha$-B$\leftrightarrow$oP8-FeB tie line. The previously unobserved oP10 structure for FeB$_4$ (SP\#58) was shown to be a high-$T$ ground state and to have the potential to be a phonon-mediated superconductor. The two proposed compact phases were shown to stabilize further under high pressures~\cite{bialon2011}. The oP10-FeB$_4$ phase has just been synthesized under medium pressures and appears to be the first realized superconductor designed entirely on the computer~\cite{Gou2013,Kolmogorov2010}. This experimental study has also led to the discovery of another new iron boride\cite{Gou2013}, Fe$_2$B$_7$, with a complex oP72 \sg{55} structure~\cite{Pettifor2013, FeStruc}, but our calculations show that the compound is metastable by about 10 meV/atom in the 0-30 GPa range. \emph{Ab initio} predictions of metastable compounds at such an unusual composition and with such a large unit cell would have been no less than an act of clairvoyance.   Nevertheless, the original DFT study~\cite{Kolmogorov2010} indicated the likely existence of new Fe-B compounds and successfully guided the experiment to discovery of new materials in a seemingly well-studied binary system.  

At 30 GPa, the only change in the set of the ground states is the stabilization of oP8-FeB over the competing oS8 and tI16 polymorphs.  After initially gaining in stability at pressures up to 20 GPa, P10-FeB$_4$ eventually becomes less stable when the tie line is defined by $\gamma$- rather than $\alpha$-B.


\subsubsection{Co-B}

Our DFT calculations show that one compound,  oP8-CoB \sg{62}~\cite{Bjurstroem1933}, is stable in the Co-B system, while references to two more additional  compounds exist in the literature.  We found tI12-Co$_2$B \sg{140}~\cite{Fruchart1959}   and the oP16-Co$_3$B \sg{62}~\cite{Bjurstroem1933}  phase to be unstable, but not far from the tie line, by 25 meV/atom and 21 meV/atom, respectively.  The oP8-CoB \sg{62} compound is also the only compound calculated to be stable at 30 GPa for Co-B.

\subsubsection{Ni-B}

The compounds in the Ni-B structures match up fairly well between those reported in literature and those calculated to be stable here.  Experimentally, both an orthorhombic and a monoclinic form of Ni$_4$B$_3$ have been seen~\cite{Rundqvist1967}, which is consistent with the calculated stability of oP28-Ni$_4$B$_3$ \sg{62} and the metastability of mS28 \sg{15} (by 2 meV/atom). \citet{Malik2013} discuss the off-stochiometric nature of oP28 and the resulting stability of mS28.  The tI12-Ni$_2$B \sg{140}~\cite{Bjurstroem1933} and the oP16-Ni$_3$B \sg{62}~\cite{Rundqvist1967} reported phases match those from the current study.  At the 1:1 composition the oS8 \sg{63} structure has been reported~\cite{Blum1952}.  According to our calculations, the three competing structures, oS8 \sg{63}, oP8 \sg{62}, and tI16 \sg{141} are all metastable with the oS8 structure being 5 meV/atom  (1meV/atom) above the tie line for GGA (LDA).   The GGA calculations indicate the possible stability of the new Ni$_5$B$_2$  compound with a mS28 \sg{15} structure (metastable by 3 meV/atom in LDA).  Therefore, this composition is worth future investigation to determine whether the GGA or LDA calculations correctly predict the structure.    Finally, \citet{Caputo2010} predicted a stable monoclinic NiB$_3$ \sg{14} structure through  computational work based upon experimental work.  As discussed in Section \ref{sec:resanal} we do not find the NiB$_3$ structure suggested by \citet{Caputo2010} to be stable nor did we find another stable structure.  At 30 GPa the newly predicted mS28-Ni$_5$B$_2$ phase is unstable, while the oS8-NiB, oP28-Ni$_4$B$_3$, tI12-Ni$_2$B, and oP16-Ni$_3$B phases remain stable.  At the Ni$_4$B$_3$ composition the metastable mS28 structure  is destabilized to 15 meV/atom above oP28.

\subsubsection{Cu-B}

The Cu-B system contains no intermediate compounds~\cite{massalski1990}.  This matches with our results at both pressures, which show all calculated structures have positive formation energies.

\subsubsection{Zn-B}

No known intermediate compounds are found for the Zn-B system~\cite{massalski1990}.  This agrees with our convex hulls for both 0 GPa and 30 GPa which shows all formation energies are positive.

\subsection{Row 4 Transition metals}

\begin{figure*}
\centering  
\includegraphics[width=0.885\textwidth]{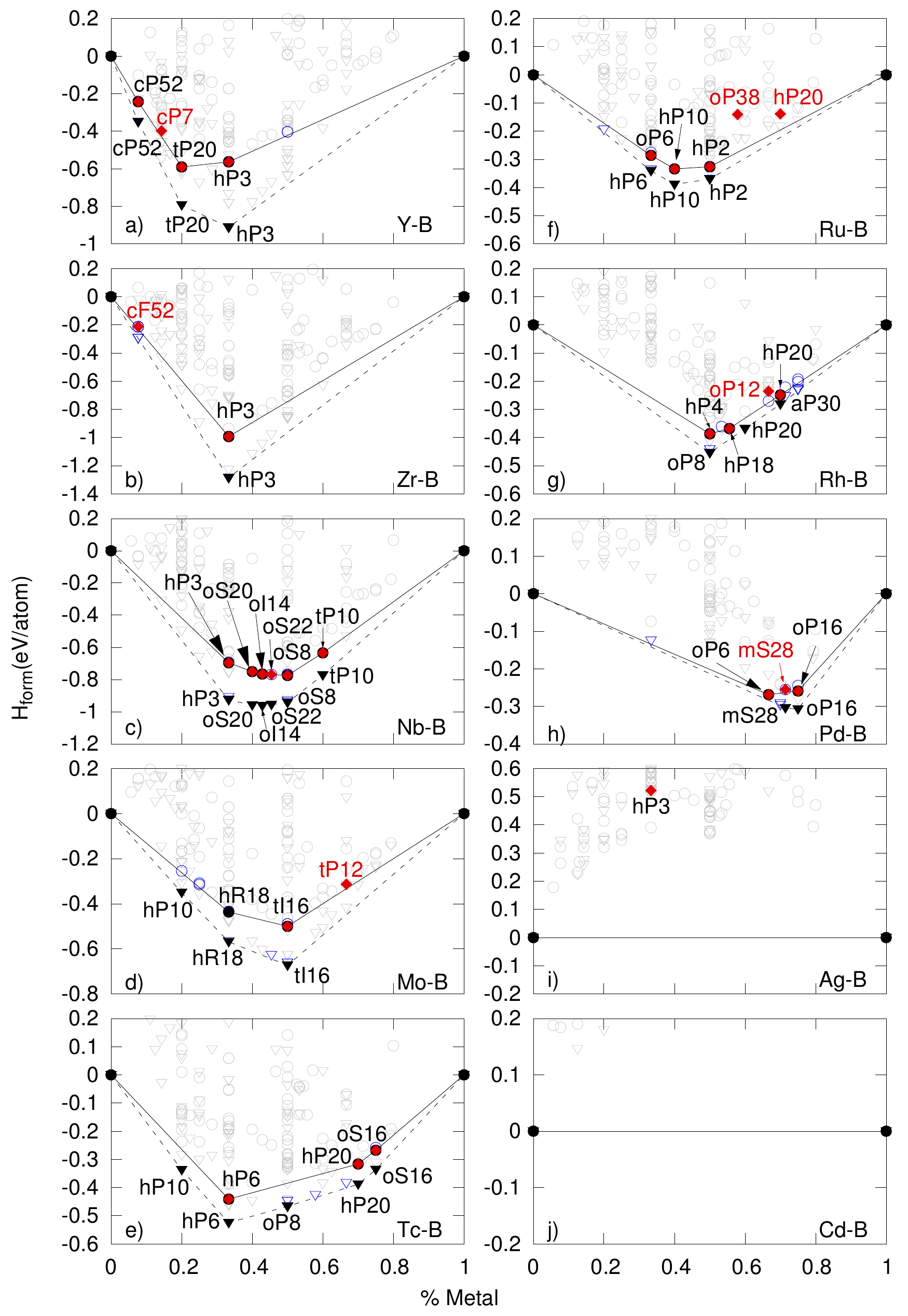}
\caption{The tie lines and convex hulls of the 4$d$ TM (Y through Cd).  Circles are 0 GPa calculations, triangles are 30 GPa calculations, and diamonds are 0 GPa reported experiments.  Black shapes are stable, gray shapes are unstable, and blue shapes are marginally-stable systems.  Black labels are stable Pearson symbols from this work, while red labels are the Pearson symbols from experiment.  If the label corresponds to both this work and literature the label is black.  For Nb the oS22 is included with a black (stable) label due to the LDA data even though the GGA data plotted shows it as metastable}\label{img:tran4}
\end{figure*}

Fig. (\ref{img:tran4}) contains the tie lines of the 4$d$ TM.

\subsubsection{Y-B}

All known compounds in the Y-B system are boron-rich. The hP3-YB$_2$ \sg{191}~\cite{Manelis1970}, tP20-YB$_4$ \sg{127}~\cite{Giese1965}, and cP7-YB$_6$ \sg{221}~\cite{Blum1954} phases have common M-B structures. Of these three phases, cP7-YB$_6$ was found to not be stable (at 31 meV/atom above the tie line) within our calculations. The set of intercalation compounds is comprised of cF52-YB$_{12}$~\cite{Matkovic1965},YB$_{25}$~\cite{Tanaka1997},YB$_{50}$~\cite{Tanaka1994}, and YB$_{66}$~\cite{Schwetz1972}. Of these complex phases, only cF52-YB$_{12}$ \sg{225}~\cite{Matkovic1965} was considered, and it was found to be stable. No additional phases are calculated to be stable at 30 GPa.

\subsubsection{Zr-B}

The Zr-B system contains two phases\cite{massalski1990}, hP3-ZrB$_2$ \sg{191}~\cite{Norton1949} and cF52-ZrB$_{12}$ \sg{225}~\cite{Matkovic1965}. The hP3-ZrB$_2$ phase is stable in our calculations, while cF52-ZrB$_{12}$ is metastable (at 18 meV/atom above the tie line). \citet{Rogl1988} discussed the metastability of both oP8-ZrB \sg{62} and oI14-Zr$_3$B$_4$ \sg{71}, which we found to be in reasonable agreement with our calculations, which are 70 meV and 43 meV above the tie line respectively.  The ground state does not change at 30 GPa.

\subsubsection{Nb-B}

The Nb-B and V-B systems contain the exact same set of stable structures noted in the V-B section to display an underlying relationship across the composition range. The reported Nb-B phases, hP3-NbB$_2$ \sg{191}, oS20-Nb$_2$B$_3$ \sg{63}, oI14-Nb$_3$B$_4$ \sg{71}, oS22-Nb$_5$B$_6$ \sg{65}, oS8-NbB \sg{63}, and tP10-Nb$_3$B$_2$ \sg{127}~\cite{Norton1949, okada1991, Andersson1950, Nowotny1958, massalski1990,Nunes2005} match our calculated ground states at 0 GPa and remain stable at 30 GPa. It is worth noting that Nb$_3$B$_2$ was reported~\cite{Zakharov1985} to possibly  be vacancy- or surface-stabilized contrary to our calculated result.

\subsubsection{Mo-B}

The Mo-B system has received a lot of recent attention due to the potential high hardness of related Mo- and W-based borides~\cite{Liang2012b,Zhang2010,Liang2012, Brazhkin2002}. \citet{Zhang2010} performed evolutionary searches at specific compositions and found that the convex hull consists of known M-B structures. After performing evolutionary searches and large scans of known M-B structures, we have obtained a matching set of the ground states. However, not all of the calculated stable structures agree with experimentally observed ones. The reported tI12-Mo$_2$B phase \sg{140} ~\cite{Kiessling1947} is not stable (at 20 meV/atom above the tie line) within our calculations or those of \citet{Zhang2010}. Theory also indicates the stability of hR18-MoB$_2$ \sg{166}~\cite{Zhang2010,Liang2012b}, while the only experimentally reported (high-temperature) structure at this composition is hP3 \sg{191}~\cite{Kiessling1947,massalski1990}. It is unlikely that temperature effects could stabilize the hP3 structure found to be over 160 meV/atom above hR18 at zero temperature. The hR21-Mo$_2$B$_5$ phase has been studied both experimentally~\cite{Kiessling1947} and theoretically~\cite{Shein2007}. According to our calculations and those of \citet{Zhang2010}, the phase is unstable by over 0.4 eV/atom and the lowest-energy structure at the Mo$_2$B$_5$ composition (24 meV/atom above the tie line) is hP14 \sg{194} suggested by \citet{Zhang2010}. These authors also found a metastable hR24 \sg{166} structure (13 meV/atom above the tie line in our calculations) with an evolutionary search for MoB$_3$ \cite{Zhang2010}, which matches the structure predicted with an evolutionary search for WB$_3$ by \citet{Liang2013}. This is not consistent with the experimental observation of the hP16-Mo$_x$B$_3$ phase \sg{194}~\cite{Lundstroem1973} that appears unstable computationally~\cite{Zhang2010,Liang2012b} (19 meV/atom above the tie line in our calculations).  One more metastable phase predicted by \citet{Zhang2010} is hP10-MoB$_4$ \sg{194} (8 meV above the tie line in our database), energetically preferred over the previously reported hP20-MoB$_4$~\cite{Galasso1968} phase, which is also dynamically unstable~\cite{Zhang2010,Liang2012b}. Finally, two polymorphs of MoB, the $\alpha$ tI16-MoB \sg{141}~\cite{Steinitz1952,massalski1990} phase and the high temperature $\beta$ oS8-MoB-\sg{64} phase~\cite{Haschke1966,Rudy1963}, have been reported, which agrees with the stability of the former and the metastability of the latter (by 11 meV/atom) in our calculations. The most recent phase diagram for Mo-B is from Ref. \cite{rogl1992}.  It includes the disputed Mo$_{1-x}$B$_3$ composition, which is discussed in more detail for the W-B system.  At 30 GPa the hR18-MoB$_2$ and tI16-MoB phases remain stable, while the previously metastable hP10-MoB$_4$ becomes stable. 
\subsubsection{Tc-B}

The  Tc-B system contains the hP6-TcB$_2$ \sg{194}, hP20-Tc$_7$B$_3$ \sg{186}, and oS16-Tc$_3$B \sg{63} ~\cite{Trzebiatowski1964} phases matching out calculated ground states.  At 30 GPa, the oP8-TcB \sg{62} phase is seen to be stable along with the hP10-TcB$_4$ \sg{194} structure of the MoB$_4$ prototype.  The ambient pressure structures remain stable as well at 30 GPa.

\subsubsection{Ru-B}
 

The most recent Ru-B phase diagram is in \citet{rogl1992}.  Among the reported hP2-RuB \sg{187}~\cite{massalski1990}, hP10-Ru$_2$B$_3$ \sg{194}~\cite{Obrowski1963,Frotscher2012}, oP6-RuB$_2$ \sg{59}~\cite{Aronsson1962}, and hP20-Ru$_7$B$_3$ \sg{186}~\cite{Aronsson1959}  phases, only the last one is found to be unstable (by 57 meV/atom) in our calculations. An oP38-Ru$_{11}$B$_8$ \sg{55}~\cite{Aselius1960,rogl1992} phase has also been reported, but does not appear in \citet{massalski1990} and is 134 meV/atom above the tie line in our database.  At 30 GPa, hP2-RuB \sg{187}, hP10-Ru$_2$B$_3$ \sg{194}, and the hP6-RuB$_2$ \sg{194} phases are stable, while hP20-Ru$_7$B$_3$ remains unstable (by 39 meV/atom).

\subsubsection{Rh-B}

The Rh-B system contains hP20-Rh$_7$B$_3$ \sg{186}~\cite{Aronsson1960}, oP12-Rh$_2$B \sg{62}~\cite{Mooney1954}, hP18-Rh$_5$B$_4$ \sg{194}~\cite{Nolaeng1981}, and hP4-RhB$_{1.1}$ \sg{194}~\cite{Aronsson1960,Aronsson1959b}.  The first two match our calculated ground states. The oP12-Rh$_2$B is found to be unstable by 70 meV/atom and the lowest-energy structure at this composition in our database is oP6 \sg{58}, a (Pd$_2$B prototype) metastable by 6 meV/atom. We constructed and simulated supercell structures of hP2 and hP4 for RhB$_x$ (0.80$<x<$1.285) and found all of them to be above the tie line, e.g. the hP15-Rh$_8$B$_7$ derivative of hP4 is metastable by 15 meV/atom. It appears that, in contrast to the case of IrB$_{0.9}$, there is no thermodynamic force for the stable hP4-RhB phase to go off-stoichiometry. At 30 GPa our calculations show that hP20-Rh$_7$B$_3$ remains stable, hP4-RhB gives way to oP8-RhB \sg{62}, and the hP18-Rh$_5$B$_4$ phase at 55\% metal content is replaced by a low-symmetry aP30-Rh$_3$B$_2$ \sg{2} phase at 60\%.

\subsubsection{Pd-B}

All observed compounds in the Pd-B systems are metal-rich: Pd$_{16}$B$_3$~\cite{Gusev2011}, Pd$_6$B\cite{Gusev2011}, Pd$_5$B\cite{Gusev2011}, oP6-Pd$_2$B \sg{58}~\cite{Beck2001,Tergenius1980}, oP16-Pd$_3$B \sg{62} \cite{Beck2001,Stenberg1961}, and mS28-Pd$_5$B$_2$ \sg{15}~\cite{Beck2001,Stenberg1961}.  \citet{Gusev2011} discusses the consensus that the three compounds with the highest Pd content are in fact disordered and ordered solid solutions of fcc Pd. For this reason, their simulation is beyond the scope of this study.  The stability of oP16-Pd$_3$B is consistent with our DFT results. The mS28-Pd$_5$B$_2$ phase is metastable in the GGA (by 8 meV/atom) but stable in the LDA (by 4 meV/atom below the tie line defined by oP6-Pd$_2$B and  oP16-Pd$_3$B). While it has been demonstrated that Pd$_2$B is amorphous~\cite{Beck2001}, the ordered oP6-Pd$_2$B \sg{62} phase reported by \citet{Tergenius1980} is stable in our calculations. According to our calculations at 30 GPa, the Pd-B system  contains only the oP16-Pd$_3$B and mS28-Pd$_5$B$_2$ stable phases.

\subsubsection{Ag-B}

There have been reports on the formation of the AgB$_2$ compound ~\cite{Obrowski1961, Islam2007,tobin2008}, but its long-term instability has been acknowledged by \citet{massalski1990} and \citet{Islam2007}. Our DFT calculations indicate the immiscibility of Ag and B at both 0 or 30 GPa. The hP3-AgB$_2$ \sg{191} phase, in particular, has a positive formation energy of 0.52 eV/atom at ambient pressure.   The lowest formation energy AgB$_2$ structure in our calculations was the hR18 \sg{166} structure, but it still retained a position formation energy.  

\subsubsection{Cd-B}

No known intermediate compounds are found for the B-Cd system~\cite{massalski1990}.  This agrees with our convex hull which shows all formation energies are positive.

\subsection{Row 5 Transition metals}

\begin{figure*}
\centering 

\includegraphics[width=0.885\textwidth]{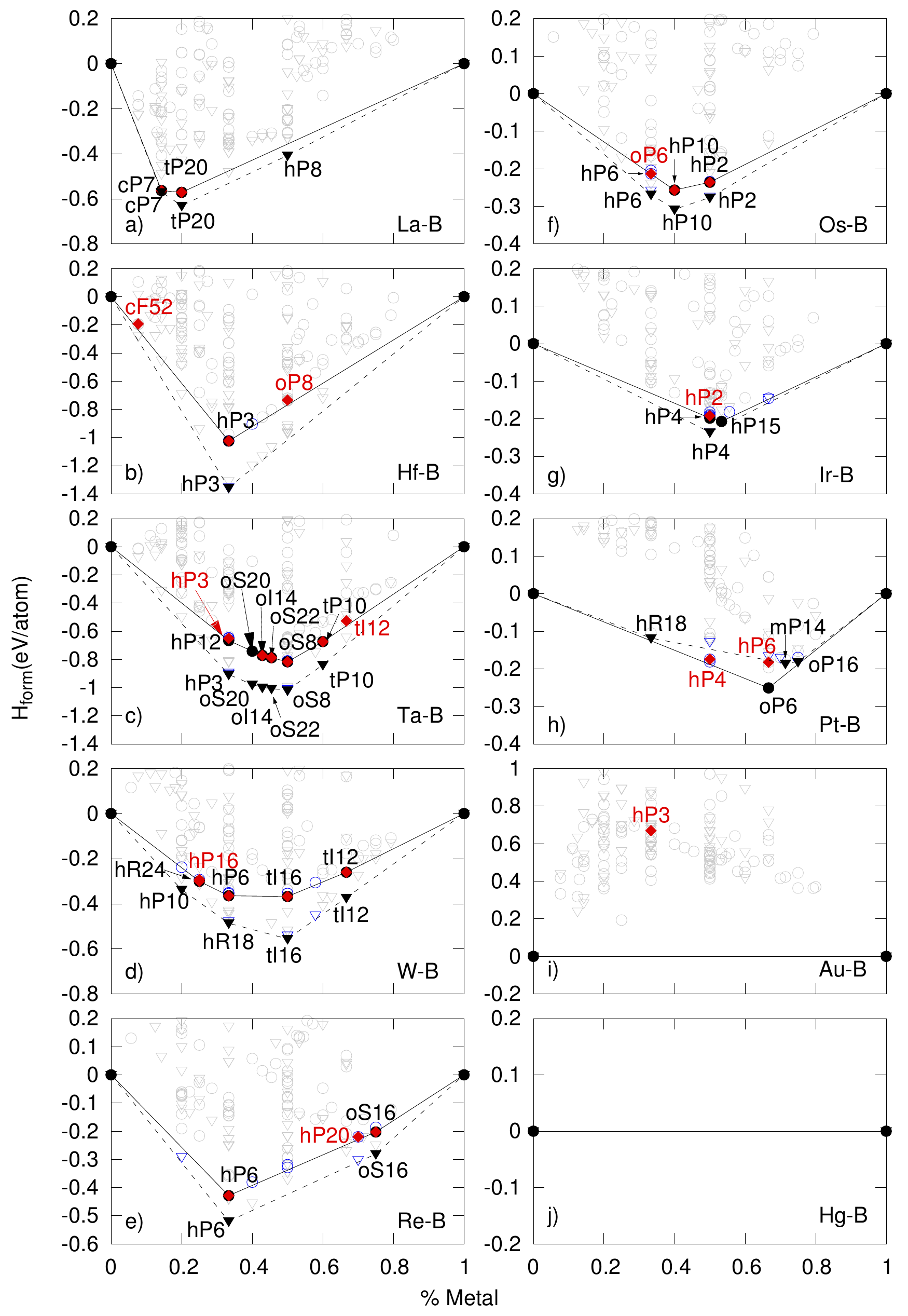}

\caption{The tie lines and convex hulls of the 5$d$ TM (La through Hg).  Circles are 0 GPa calculations, triangles are 30 GPa calculations, and diamonds are 0 GPa reported experiments.  Black shapes are stable, gray shapes are unstable, and blue shapes are marginally-stable  systems.  Black labels are stable Pearson symbols from this work, while red labels are the Pearson symbols from experiment.  If the label corresponds to both this work and literature the label is black. For Os-B the hP6 for ambient pressures is stable in LDA and not GGA.}\label{img:tran5}
\end{figure*}

The 5$d$ transitions metals are discussed below with Fig. (\ref{img:tran5}) containing the corresponding phase diagrams.

\subsubsection{La-B}

The La-B system contains the cP7-LaB$_6$ \sg{221}~\cite{Eliseev1986} and tP20-LaB$_4$ \sg{127} phases~\cite{Kato1974}, which agrees with this study.  At 30 GPa, an additional hP8-LaB \sg{194} phase defines the convex hull. Among the considered binaries, the predicted hP8  (metal sandwich) structure~\cite{Kolmogorov2006b,Kolmogorov2006} is found to become stable only for LiB and LaB in the considered pressure range.

\subsubsection{Hf-B}

 Several compounds have been reportedly observed in the Hf-B system: cF52-HfB$_{12}$ \sg{225}, hP3-HfB$_2$ \sg{191},  cF8-HfB \sg{225}, and oP8-HfB \sg{53}~\cite{Cannon1983,Post1954,Glaser1953b, Rogl1988Hf}. However, \citet{Rogl1988Hf} demonstrated that cF8-HfB is C stabilized, while oP8-HfB is stable at high temperatures. Our calculations show that only hP3-HfB$_2$ is stable. The cF52-HfB$_{12}$ and cF8-HfB phases are above the tie line at 44 meV/atom and 350 meV/atom, respectively. The lowest-energy structure at the 1:1 composition in our database, oP8-HfB \sg{62}, is still 34 meV/atom above the tie line agreeing with experiment~\cite{Rogl1988Hf} that neither cF8 nor oP8 are stable at low temperature. hP3-HfB$_2$ is the only calculated stable phase at 30 GPa as well.

\subsubsection{Ta-B}

The oS8-TaB \sg{63}~\cite{Okada1993}, oS22-Ta$_5$B$_6$ \sg{65}~\cite{Bolmgren1990,Okada1993}, oI14-Ta$_3$B$_4$ \sg{71}~\cite{Kiessling1949, Okada1993} and tP10-Ta$_3$B$_2$ \sg{127}~\cite{Nowotny1958} phases reported  in the literature agree with our database. However, we have found several discrepancies. For the TaB$_2$ composition hP3-TaB$_2$ \sg{191} has been observed experimentally~\cite{Norton1949,Okada1993}. The employed DFT approximations order the known competing structure types differently. The GGA favors the hP12 \sg{194} structure over hR18 \sg{166} by 6 meV/atom and over hP3 \sg{191} by 15 meV/atom, while the LDA favors the hP3\sg{191} structure over hP12 \sg{194} by 12 meV/atom and over hR18 \sg{166} by 18 meV/atom. The oS20-Ta$_2$B$_3$ \sg{63} phase is calculated to be marginally stable (2 meV/atom below the tie line defined by hP12-TaB$_2$ and oI14-Ta$_3$B$_4$), but has not been reported experimentally.  See Section \ref{sec:resanal} for further details. Finally, the tI12-Ta$_2$B \sg{140} reported phase~\cite{Havinga1972} is not stable in our calculations at 35 meV above the tie line, which is consistent with \citet{Chad2006} who found it to be a high temperature phase.  At 30 GPa the stable Ta-B structures do not change from their ambient state except that hP3-TaB$_2$ is found to be stable with both functionals.

\subsubsection{W-B}


The W-rich portion of the W-B system is well established in literature with tI12-W$_2$B~\cite{ Zhao2010,Liang2011,Aronsson1968} and tI16-WB~\cite{Zhao2010,Liang2011,Kiessling1947} reported both experimentally and theoretically. A $\beta$ phase of WB, oS8 \sg{63}, has also been reported~\cite{Haschke1966}. The calculated stability of tI12-W$_2$B \sg{140} and tI16-WB \sg{141} as well as the metastability of oS8-WB \sg{63} (by 13 meV/atom) are consistent with these experimental observations. The B-rich phases show less agreement between theory and experiment. For the WB$_2$ composition, experiments indicate formation of the hP12 \sg{194} structure~\cite{Aronsson1968}, while alternative hP6 \sg{194}~\cite{Zhao2010,Liang2011} and oP6 \sg{59}~\cite{Chen2008} solutions have been proposed as lowest-energy structures theoretically. In our calculations, hP6 is stable while the oP6 and hP12 structures are metastable by 9 meV/atom and 30 meV/atom, respectively. It is worth noting that the previously reported W$_2$B$_5$ composition is now viewed to be the hP12-WB$_2$ phase~\cite{Zhao2010}. A debate has recently sparked regarding the true composition and structure of a known (possibly super hard)~\cite{Liang2012, Brazhkin2002} boron-rich W-B compound, with theory (this study included) predicting hP16-WB$_3$ \sg{194} and experiment finding hP20-WB$_4$ \sg{194}~\cite{Zhao2010, Liang2012, Liang2011, Xie2012} or hP-W$_{1-x}$B$_3$ ~\cite{Duschanek1995} (where  x$\sim$0.2 and is isotypic with hP20-WB$_4$). The hP16-WB$_3$ phase can be morphed into hP20-WB$_4$ by inserting a B dimer that converts the 2D boron network into the 3D one.  \citet{Zhang2012} have demonstrated that the structures' similarity has caused confusion in the analysis of experimental XRD~\cite{Mohammadi2012,Mohammadi2011} data due to all of the WB$_3$ XRD peaks existing in the WB$_4$ data. In addition, the high pressure stability of WB$_4$~\cite{Zhao2010} obfuscates comparison of the DFT and experimental results. Finally, \citet{Liang2013} have proposed an additional WB$_3$ prototype, hR24 \sg{166}, which is expected to be slightly more energetically favorable than the previously known hP16 structure at zero temperature, agreeing with our calculations. \citet{Cheng2013} showed that inclusion of the vibrational entropy term to the free energy destabilizes hR24 with respect to hP16 at 660-680$^\circ$C. Then experimentally and theoretically, they demonstrated the existence of a WB$_{3+x}$ phase that contains B interstitials in the W plane voids.  Within our calculations we considered both B interstitials (single and double) and W vacancies within the hP16-WB$_3$ phase to understand the hP-W$_{1-x}$B$_3$ and hP-WB$_{3+x}$ phases and saw that both defects increased the formation energy within a single conventional cell.

At 30 GPa the calculated convex hull changes to hP10-WB$_4$, hR18-WB$_2$, tI16-WB, and tP6-W$_2$B.  This contradicts the study of \citet{Xie2012} who compressed WB$_4$ up to 60 GPa demonstrating a structural change at 42 GPa with the structure not changing phase before this transition.

\subsubsection{Re-B}

The Re$_3$B, Re$_7$B$_3$, ReB$_3$, and ReB$_2$ compositions have been reported\cite{massalski1990, Aronsson1960b, Aronsson1960, Placa1962}, with  ReB$_3$  suggested theoretically as unstable at ambient conditions due to a positive formation energy~\cite{Zhao2010a}. Additionally, the Re$_2$B, ReB, Re$_2$B$_3$, Re$_2$B$_5$, and ReB$_4$ compositions have all been studied with computational techniques\cite{Zhao2010a, Zhao2009, Ivanovskii2012, Soto2008} and are generally considered unstable or metastable\cite{Zhao2010a}.  Therefore, from the experimental and theoretical studies it appears that oS16-Re$_3$B  {\sg{53}, hP20-Re$_7$B$_3$ \sg{186},  and hP6-ReB$_2$  \sg{194}are the stable phases for the Re-B system.  We calculate hP6-ReB$_2$ and oS16-Re$_3$B to be stable structures with the hP20-Re$_7$B$_3$ structure being metastable (12 meV/atom above the tie line).  A metastable hP6-ReB \sg{156} structure (at 10 meV/Atom above the tie line) consisting of alternating buckled hexagonal and closed packed B layers was found using evolutionary search.  Further hP6-ReB$_2$ and oS16-Re$_3$B are stable at 30GPa in GGA and LDA. 

\subsubsection{Os-B}

The Os-B system has three reported phases, hP2-OsB \sg{51}, hP10-Os$_2$B$_3$ \sg{194}, and oP6-OsB$_2$ \sg{59}~\cite{Frotscher2012,Aronsson1962}. At the OsB$_2$ composition, the oP6 \sg{59} structure~\cite{Aronsson1962} is found to be 9 meV/atom above the hP6 \sg{194} structure that effectively lies on the $\alpha$-B$\leftrightarrow$hP10-Os$_2$B$_3$ tie line. The hP10-Os$_2$B$_3$ phase was studied by neutron diffraction by \citet{Frotscher2012} and demonstrated to have an approximate composition of OsB$_{1.6}$. The stoichiometric structure is found to be stable in our calculations.  The obtained stability of the hP2-OsB phase is in agreement with both the experimental~\cite{massalski1990} and previous \emph{ab initio} studies~\cite{Chen2011} of $\alpha$-OsB. The $\beta$-OsB phase\cite{massalski1990} has only (cubic) cell parameters reported which does not match any phases with negative formation energies in our database. The 30-GPa convex hull is defined by the same set of phases.

\subsubsection{Ir-B}


Several off-stoichiometric Ir-B phases near 1:1 composition have been reported and are shown in the most recent phase diagram~\cite{Rogl1998}: hP2-IrB$_{0.9}$ \sg{187} (often referred to as $\beta$-IrB), oS16-IrB$_{0.9}$ \sg{36}  (often referred to as $\alpha$-IrB), tI12-IrB$_{1.1}$ \sg{141}, and mS-IrB$_{1.26-1.5}$ \sg{12}~\cite{Rogl1971, Aronsson1960, Aronsson1963, Aronsson1962, massalski1990,Rogl1998}. The last three have been suggested to be high-temperature phases \cite{massalski1990}. To investigate the stoichiometry-dependent stabilization effect we simulated IrB$_x$ (0.875$<x<$1.25) phases by creating B or Ir vacancies in hP2-IrB \sg{187}  supercells. The Ir-rich hP15-Ir$_8$B$_7$ phase was indeed found to be stable rendering possible Ir$_2$B candidate compounds metastable. Although hP4-IrB still appears below the $\alpha$-B$\leftrightarrow$ hP15-Ir$_8$B$_7$ tie line, a more systematic simulation of the off-stoichiometric compounds may lead to a different outcome. The report on synthesized Ir$_3$B$_2$ contained no structural information~\cite{massalski1990} and we have not observed a stable phase at this composition among the considered prototypes. Application of pressure has been found to destabilize the constructed hP15 off-stoichiometric phase and only hP4-IrB is seen to be stable at 30 GPa.

\subsubsection{Pt-B}


Our findings raise questions about all three reported compounds in the Pt-B system~\cite{massalski1990}. No structural model is available for the tetragonal Pt$_3$B phase~\cite{massalski1990}, and the only relevant structure at this composition in our calculations is oP16 \sg{62}, Pd$_3$B-type) that is metastable by 18 meV/atom. The reported hP4-PtB  \sg{194}~\cite{Aronsson1960,massalski1990} phase with linear boron chains was considered in our previous study and noted to be in poor agreement with experiment\cite{Kolmogorov2006}. The exceptionally high B-B bond length mismatch of over 10\% observed previously \cite{Kolmogorov2006} and the positive formation energy of 55 meV/atom calculated presently call for a more detailed study of this compound. Our evolutionary search identified an unrelated hP4-PtB \sg{164} phase, just 7 meV/atom above the tie line, comprised of buckled hexagonal sheets of boron (Wyckoff positions are in the Supplementary Materials~\cite{SupMat}).  Finally, the observed Pt$_2$B phase is reported to be hP6 \sg{194} ~\cite{massalski1990}, which is found in our calculations to be 68 meV/atom above the oP6 \sg{58}, Pd$_2$B-type) structure.  An orthorhombic PtB$_{0.67}$ has also been observed experimentally. Simulation of this disordered phase is outside the scope of our study.   The only stable phase, oP6-Pt$_2$B, obtained in our calculations at 0 GPa is replaced by a set of hR18-PtB$_2$, mP14-Pt$_5$B$_2$, and oP16-Pt$_3$B phases at 30 GPa.  Interestingly, Pt-B is found to be the only considered M-B binary for which the formation enthalpy reduces with pressure. Although pressure-induced disproportionation has been seen in some systems~\cite{Williams1987,Skelton1983} we have not been able to identify factor(s) that make this system special.

\subsubsection{Au-B}

The hP3-AuB$_2$ \sg{191} phase was first reported by \citet{Obrowski1961}, but \citet{massalski1990} describe it as a metastable. In a previous study~\cite{Kolmogorov2006} the authors showed that hP3-AuB$_2$ not only has a large positive formation energy, but is also unstable with respect to a proposed lower-symmetry $\delta$-AuB$_2$ derivative. All the considered phases have been found to have positive formation energies at both 0- and 30-GPa pressures.

\subsubsection{Hg-B}

The Hg-B system contains no stable compounds~\cite{massalski1990}, which is consistent with our calculations at both pressures considered.

\begin{longtable}[t]{ l r r r r  r}
\hline\hline
 formula &\multicolumn{2}{ c }{H$_f$ (eV/atom)}&\multicolumn{2}{ c }{$dH$ (meV/atom)}&\multicolumn{1}{ c }{Pearson }\\
&GGA&LDA&GGA&LDA&  Sym. \\   
\hline 

 Li$_{3}$B$_{14}$ &  -0.2326 &  -0.2453 &0.0 &0.0 & tP136 \\
 LiB$_{3}$ &  -0.2378 &  -0.2313 &0.0 &6.6 & tP16 \\
 Li$_{8}$B$_{7}$ &  -0.1907 &  -0.2093 &0.0 &0.0 & hP15\\
 Be$_{29}$B$_{81}$ &  -0.1240 &  -0.1320 &0.0 &0.0 & hP110\\
 Na$_{3}$B$_{20}$ &  -0.0594 &  -0.0305 &0.0 &0.0 & oS46 \\
 MgB$_{7}$ &  -0.1317 &  -0.1444 &0.0 &0.0 & oI64 \\
 MgB$_{4}$ &  -0.1402 &  -0.1490 &0.0 &0.0 & oP20 \\
 MgB$_{2}$ &  -0.1321 &  -0.1496 &0.0 &0.0 & oS12 \\
 AlB$_{2}$ &  -0.0447 &  -0.0756 &0.0 &0.0 & hP3 \\
 KB$_{6}$ &  -0.0162 &  -0.0072 &0.0 &13.5 & cP7 \\
 CaB$_{6}$ &  -0.4122 &  -0.4098 &0.0 &0.0 & cP7 \\
 CaB$_{4}$ &  -0.3997 &  -0.4381 &0.0 &0.0 & tP20 \\
 ScB$_{12}$ &  -0.2105 &  -0.2167 &0.0 &0.0 & cF52 \\
 ScB$_{2}$ &  -0.8376 &  -0.9098 &0.0 &0.0 & hP3 \\
 TiB$_{2}$ &  -1.0600 &  -1.1425 &0.0 &0.4 & hP3\\
 Ti$_{3}$B$_{4}$ &  -0.9357 & -1.0002 &0.0 &0.0 & oI14 \\
 TiB &  -0.8358 &  -0.8865 &0.0 &0.0 & oP8 \\
 VB$_{2}$ &  -0.7402 & -0.7999 &0.0 &0.0 & hP3 \\
 V$_{2}$B$_{3}$ &  -0.7990 &  -0.8535 &0.0 &0.0 & oS20\\
 V$_{3}$B$_{4}$ &  -0.8202 & -0.8708 &0.0 &0.0 & oI14 \\
 V$_{5}$B$_{6}$ &  -0.8312 &  -0.8796 &0.0 &0.0 & oS22\\
 VB &  -0.8497 &  -0.8935 &0.0 &0.0 & oS8 \\
 V$_{3}$B$_{2}$ &  -0.7249 &  -0.7604 &0.0 &0.0 & tP10 \\
 CrB$_{4}$ &  -0.3098 &  -0.3533 &0.0 &0.0 & oP10\\
 CrB$_{2}$ &  -0.4206 &  -0.4639 &0.0 &0.0 & hP12 \\
 CrB &  -0.5321 &  -0.5819 &0.0 &0.0 & tI16\\
 Cr$_{5}$B$_{3}$ &  -0.4570 &  -0.7147 &0.0 &0.0 & tI32 \\
 Cr$_{2}$B &  -0.3753 &  -0.4109 &0.0 &0.0 & oF48 \\
 MnB$_{4}$ &  -0.2971 &  -0.3460 &0.0 &0.0 & mP20 \\
 MnB &  -0.5184 &  -0.5031 &0.0 &0.0 & oP8 \\
 Mn$_{2}$B &  -0.4338 &  -0.4935 &0.0 &0.0 & oF48 \\\
 FeB$_{2}$ &  -0.3001 &  -0.3877 &0.0 &0.0 & oP12 \\
 FeB &  -0.3802 &  -0.4084 &0.0 &14.6 & tI16 \\
 Fe$_{2}$B &  -0.3152 &  -0.3096 &0.0 &0.0 & tI12 \\
 CoB &  -0.4006 &  -0.5287 &0.0 &0.0 & oP8 \\
 Ni$_{4}$B$_{3}$ &  -0.2818 &  -0.3532 &0.0 &0.0 & oP28 \\
 Ni$_{2}$B &  -0.2916 &  -0.3636 &0.0 &0.0 & tI12 \\
 Ni$_{5}$B$_{2}$ &  -0.2800 &  -0.3331 &0.0 &3.3 & mS28 \\
 Ni$_{3}$B &  -0.2640 &  -0.3160 &0.0 &0.0 & oP16 \\
 SrB$_{6}$ &  -0.4510 &  -0.4604 &0.0 &0.1 & cP7 \\
 YB$_{12}$ &  -0.2432 &  -0.2555 &0.0 &0.0 & cF52 \\
 YB$_{4}$ &  -0.5899 &  -0.6266 &0.0 &0.0 & tP20 \\
 YB$_{2}$ &  -0.5633 & -0.5633 &0.0 &0.0 & hP3 \\
 ZrB$_{2}$ &  -0.9928 &  -1.0579 &0.0 &0.0 & oS6 \\
 NbB$_{2}$ &  -0.6961 &  -0.7456 &0.0 &0.0 & hP3 \\
 Nb$_{2}$B$_{3}$ &  -0.7507 &  -0.7909 &0.0 &0.0 & oS20 \\
 Nb$_{3}$B$_{4}$ &  -0.7659 & 0.8013 &0.0 &0.0 & oI14 \\
 NbB &  -0.7745 &  -0.7991 &0.0 &0.0 & oS8 \\
 Nb$_{3}$B$_{2}$ &  -0.6346 &  -0.6546 &0.0 &0.0 & tP10\\
 MoB$_{2}$ &  -0.4366 &  -0.4726 &0.0 &0.0 & hR18 \\
 MoB &  -0.5002 &  -0.5384 &0.0 &0.0 & tI16 \\
 TcB$_{2}$ &  -0.4409 &  -0.4673 &0.0 &0.0 & hP6\\
 Tc$_{7}$B$_{3}$ &  -0.3164 &  -0.3237 &0.0 &0.0 & hP20\\
 Tc$_{3}$B &  -0.2677 &  -0.2764 &0.0 &0.0 & oS16 \\
 RuB$_{2}$ &  -0.2859 &  -0.3033 &0.0 &0.0 & oP6 \\
 Ru$_{2}$B$_{3}$ &  -0.3337 &  -0.3569 &0.0 &0.0 & hP10\\
 RuB &  -0.3260 &  -0.3471 &0.0 &0.0 & hP2 \\
 RhB &  -0.3868 &  -0.4108 &0.0 &0.0 & oS8 \\
 Rh$_{5}$B$_{4}$ &  -0.3680 &  -0.3782 &0.0 &0.0 & hP18 \\
 Rh$_{7}$B$_{3}$ &  -0.2488 &  -0.2499 &0.0 &5.4 & hP20 \\
 Pd$_{2}$B &  -0.2687 &  -0.2838 &0.0 &0.0 & oP6 \\
 Pd$_{3}$B &  -0.2590 &  -0.2913 &0.0 &0.0 & oP16 \\
 BaB$_{6}$ &  -0.4096 &  -0.4251 &0.0 &0.0 & cP7 \\
 LaB$_{6}$ &  -0.5637 &  -0.5710 &0.0 &0.0 & cP7 \\
 LaB$_{4}$ &  -0.5713 &  -0.5877 &0.0 &0.0 & tP20 \\
 HfB$_{2}$ &  -1.0244 &  -1.1097 &0.0 &0.0 & hP3 \\
 TaB$_{2}$ &  -0.6658 &  -0.7004 &0.0 &12.3 & hP12 \\
 Ta$_{2}$B$_{3}$ &  -0.7417 &  -0.7944 &0.0 &0.0 & oS20\\
 Ta$_{3}$B$_{4}$ &  -0.7717 & -0.8197 &0.0 &0.0 & oI14 \\
 Ta$_{5}$B$_{6}$ &  -0.7884 &  -0.8335 &0.0 &0.0 & oS22\\
 TaB &  -0.8168 &  -0.8544 &0.0 &0.0 & oS8 \\
 Ta$_{3}$B$_{2}$ &  -0.6734 &  -0.7048 &0.0 &0.0 & tP10\\
 WB$_{3}$ &  -0.3001 &  -0.3194 &0.0 &0.0 & hR24 \\
 WB$_{2}$ &  -0.3644 &  -0.3854 &0.0 &0.0 & hP6 \\
 WB &  -0.3679 &  -0.4150 &0.0 &0.0 & tI16 \\
 W$_{2}$B &  -0.2602 &  -0.2828 &0.0 &0.0 & tI12\\
 ReB$_{2}$ &  -0.4287 &  -0.4549 &0.0 &0.0 & hP6 \\
 Re$_{3}$B &  -0.2028 &  -0.2148 &0.0 &0.0 & oS16 \\
 Os$_{2}$B$_{3}$ &  -0.2567 &  -0.2743 &0.0 &0.0 & hP10\\
 OsB &  -0.2359 &  -0.2496 &0.0 &0.0 & hP2 \\
 IrB &  -0.1982 &  -0.2008 &0.0 &4.3 & hP4 \\
 Ir$_{8}$B$_{7}$ &  -0.2071 &  0.2058 &0.0 &0.0 & hP15\\
 Pt$_{2}$B &  -0.2509 &  -0.2379 &0.0 &0.0 & oP6 \\
   \hline\hline
    \caption{The compounds calculated to be stable within GGA at ambient pressures with the corresponding $dH$ of the LDA calculations. Here $dH$ is the stability in relation to phase separation as defined by distance of H$_f$ to the tie line. Compounds with $dH$=0 are ground states. Compounds with 0$<dH<$20 meV/atom are termed metastable (only for LDA). A 20 meV/atom cutoff, a typical size of the \emph{relative} contribution to the Gibbs energy from the vibrational entropy term at $T\sim1000$ K~\cite{Kolmogorov2010}, is chosen to distinguish between possible metastable and unstable phases. Compounds with $dH$ values greater than 20 meV/atom are not included here.}\label{tab:stab}
\end{longtable}

\begin{longtable}{ l r r r r r r }
\hline\hline
 formula &\multicolumn{2}{ c }{H$_f$ (eV/atom)}&\multicolumn{2}{ c }{$dH$ (meV/atom)}&\multicolumn{1}{ c }{Pearson}\\
&GGA&LDA&GGA&LDA&Sym.\\
\hline 

 Li$_{10}$B$_{9}$ &  -0.1913 &  -0.2058 &0.6 &4.1 & hP19  \\
 Li$_{6}$B$_{5}$ &  -0.1808 &  -0.2064 &5.0 &0.0 & hP11  \\
 Be$_{3}$B$_{50}$ &  -0.0255 &  -0.0173 &1.1 &11.0 & aP53  \\
 BeB$_{2}$ &  -0.0994 &  --- &12.8 &--- & oP12\\
 BeB$_{2}$ &  -0.0991 &  --- &13.1 &--- & oS12  \\
 Be$_{4}$B &  -0.0165 &  -0.0211 &17.2 &14.8 & tP10 \\
 NaB$_{6}$ &  -0.0393 &  -0.0258 &19.3 &7.1 & oS28 \\
 NaB$_{3}$ &  -0.0448 &  -0.0537 &6.4 &0.0 & tP16 \\
 MgB$_{2}$ &  -0.1302 &  -0.1479 &1.9 &1.8 & hP3 \\
 MgB$_{2}$ &  -0.1313 &  -0.1486 &0.8 &1.0 & oP12 \\
 MgB &  -0.0892 &  -0.1444 &9.9 &0.0 & hP8 \\
 MgB &  -0.0906 &  -0.1087 &8.4 &3.5 & hR4 \\
 Mg$_{3}$B$_{2}$ &  -0.0608 &  -0.0771 &18.5 &12.6 & hP5\\
 K$_{3}$B$_{20}$ &  -0.0113 &  -0.0209 &3.5 &0.0 & oS46 \\
 ScB &  -0.6122 &  --- &16.0 &--- & hP8 \\
 ScB &  -0.6146 &  -0.6634 &13.6 &18.9 & hR4 \\
 Ti$_{2}$B$_{3}$ &  -0.9712 &  -1.0414 &1.8 &1.7 & oS20 \\
 Ti$_{5}$B$_{6}$ &  -0.8961 &  -0.9564 &3.3 &2.5 & oS22 \\
 TiB &  -0.8306 &  -0.8826 &5.1 &3.9 & oS8 \\
 VB$_{2}$ &  -0.7270 &  -0.7870 &13.2 &12.9 & tI12 \\
 VB &  -0.8409 &  -0.8826 &8.8 &10.9 & oP8 \\
 VB &  -0.8380 &  -0.8824 &11.7 &11.1 & tI16 \\
 VB &  -0.8406 &  -0.8826 &9.1 &10.9 & oP8 \\
 CrB$_{4}$ &  -0.3043 &  -0.4276 &5.5 &8.2 & mS20 \\
 Cr$_{5}$B$_{6}$ &  -0.4854 &  -0.6013 &16.2 &0.0 & oS22 \\
 CrB &  -0.5224 &  -0.7773 &9.6 &10.8 & oS8\\ 
 Cr$_{3}$B$_{2}$ &  -0.4301 &  -0.7223 &11.2 &7.0 & tP10\\
 Cr$_{2}$B &  -0.3744 &  -0.6852 &1.0 &0.6 & tI12 \\
  MnB$_{2}$ &  -0.3763 &  -0.4256 &19.2 &0.0 & hP6 \\
   MnB$_{4}$ &  -0.2791 &  --- &18.0 &--- & oI10 \\
 MnB$_{4}$ &  -0.2888 &  -0.5276 &8.2 &18.1 & oP10 \\
 MnB$_{4}$ &  -0.2775 &  --- &19.6 &--- & mS5 \\
 MnB$_{4}$ &  -0.2857 &  -0.5316 &11.4 &14.1 & mS20 \\
 MnB$_{4}$ &  -0.2954 &  -0.5437 &1.6 &2.0 & oF80 \\
 MnB &  -0.5000 &  -1.0019 &18.3 &0.3 & tI16 \\
 MnB &  -0.5111 &  --- &7.3 &--- & oS8 \\
 MnB &  -0.5115 &  -0.9952 &6.9 &7.1 & oS8 \\
 Mn$_{2}$B &  -0.4286 &  -1.1549 &5.1 &4.1 & tI12 \\
 Mn$_{3}$B &  -0.3063 &  -1.1590 &19.1 &0.0 & oP16 \\
 Mn$_{23}$B$_{6}$ &  -0.2518 &  -1.0630 &17.4 &0.0 & cF29 \\
 FeB$_{4}$ &  -0.1698 &  -0.2218 &10.2 &10.8 & oP10 \\
 FeB$_{4}$ &  -0.1635 &  --- &16.6 &--- & mS30 \\
 Fe$_{2}$B$_{7}$ &  -0.1895 &  -0.2414 &10.6 &17.0 & oP72 \\
 FeB &  -0.3746 &  -0.4187 &5.6 &4.4 & oP8  \\
 FeB &  -0.3747 &  -0.4230 &5.5 &0.0 & oS8 \\
 Fe$_{3}$B &  -0.2183 &  --- &18.1 &--- & oP16\\
 Fe$_{3}$B &  -0.2169 &  --- &19.5 &--- & oS16 \\
 Fe$_{23}$B$_{6}$ &  -0.1758 &  --- &19.8 &--- & cF116\\
 NiB &  -0.2386 &  -0.3044 &8.0 &4.6 & oP8 \\
 NiB &  -0.2282 &  -0.2934 &18.3 &15.6 & tI16\\
 NiB &  -0.2414 &  -0.3085 &5.2 &0.5 & oS8 & \\
 Ni$_{4}$B$_{3}$ &  -0.2796 &  -0.3459 &2.2 &7.3 & mS14 \\
 Ni$_{3}$B$_{2}$ &  -0.2673 &  --- &17.4 &--- & aP30\\
 Ni$_{7}$B$_{3}$ &  -0.2695 &  -0.3252 &14.0 &19.3 & hP20\\
 YB &  -0.4032 &  --- &19.3 &--- & oS8\\
 ZrB$_{12}$ &  -0.2147 &  --- &14.4 &--- & tI26 \\
 Nb$_{5}$B$_{6}$ &   -0.7689 &  -0.8008 & 0.2 & 0.0 &  oS22 \\
 NbB &  -0.7681 &  -0.6546 &6.4 &0.0 & oP8 \\
 MoB$_{4}$ &  -0.2541 &  -0.2812 &7.8 &2.4 & hP10\\
 MoB$_{3}$ &  -0.3086 &  --- &18.9 &--- & hP16\\
 MoB$_{3}$ &  -0.3144 &  --- &13.0 &--- & hR24 \\
 MoB$_{2}$ &  -0.4338 &  -0.4693 &2.7 &3.3 & hP12\\
 MoB &  -0.4892 &  -0.5265 &11.1 &11.9 & oS8\\
 Tc$_{3}$B &  -0.2582 &  -0.2610 &9.4 &15.4 & oP16\\
 RuB$_{2}$ &  -0.2759 &  -0.2991 &10.0 &4.2 & hP6 \\
 Rh$_{8}$B$_{7}$ &  -0.3609 &  --- &14.6 &--- & hP15\\
 Rh$_{2}$B &  -0.2702 &  --- &6.0 &--- & oP6 \\
 Rh$_{5}$B$_{2}$ &  -0.2215 &  -0.2234 &15.4 &19.7 & mS28 \\
 Rh$_{3}$B &  -0.1928 &  -0.1942 &14.5 &18.6 & oP16 \\
 Rh$_{3}$B &  -0.2018 &  -0.2010 &5.5 &11.7 & oS16 \\
 Pd$_{5}$B$_{2}$ &  -0.2554 &  -0.2926 &7.8 &0.0 & mS28\\
 Pd$_{3}$B &  -0.2444 &  -0.2717 &14.6 &19.7 & oS16\\
 Hf$_{2}$B$_{3}$ &  -0.9031 &  --- &18.9 &--- & oS20\\
 TaB$_{2}$ &  -0.6508 &  --- &15.0 &0.0 & hP3\\
 TaB$_{2}$ &  -0.6599 &  -0.6943 &5.9 &18.3 & hR18\\
 TaB$_{2}$ &  -0.6468 &  -0.7093 &19.0 &3.3 & oP12 \\
 TaB &  -0.8114 &  -0.8482 &5.4 &6.2 & oP8 \\
 WB$_{4}$ &  -0.2380 &  -0.2673 &2.1 &0.0 & hP10\\
 WB$_{3}$ &  -0.2923 &  -0.3109 &7.8 &8.5 & hP16 \\
 WB$_{2}$ &  -0.3555 &  -0.3773 &8.9 &8.1 & oP6 \\
 WB$_{2}$ &  -0.3512 &  -0.3719 &13.2 &13.5 & hR9\\
 WB &  -0.3541 &  -0.4005 &13.8 &14.5 & oS8\\
 W$_{11}$B$_{8}$ &  -0.3056 &  -0.3408 &11.3 &11.6 & oP38\\
 Re$_{2}$B$_{3}$ &  -0.3806 &  -0.4026 &12.0 &13.9 & hP10 \\
 ReB &  -0.3187 &  --- &19.6 &--- & tI8\\
 ReB &  -0.3283 &  -0.3448 &10.0 &14.1 & hP6 \\
 Re$_{7}$B$_{3}$ &  -0.2201 &  -0.2313 &9.8 &12.3 & hP20 \\
 Re$_{3}$B &  -0.1865 &  --- &16.3 &--- & oP16\\
 OsB$_{2}$ &  -0.2045 &  -0.2240 &9.4 &4.6 & hP6 \\
OsB$_{2}$ &  -0.2134 &-0.2261 &0.5 &2.5 & oP6\\
 IrB &  -0.1825 &  --- &15.8 &--- & tP4\\
 IrB &  -0.1918 &  -0.1967 &6.4 &8.5 & hP2\\
 IrB &  -0.1945 &  -0.2052 &3.7 &0.0 & tI8\\
 Ir$_{5}$B$_{4}$ &  -0.1817 &  -0.1764 &15.6 &6.0 & hP18\\
 Ir$_{2}$B &  -0.1454 &  --- &2.6 &--- & hP6\\
 PtB &  -0.1747 &  -0.1764 &13.5 &10.4 & hR4\\
 PtB &  -0.1815 &  -0.1868 &6.7 &0.0 & hP4 \\
 Pt$_{3}$B &  -0.1699 &  -0.1713 &18.2 &7.1 & oP16\\

   \hline\hline
\caption{The compounds calculated to be metastable at ambient pressures within GGA with the corresponding $dH$ from LDA.  Here $dH$ is the stability in relation to phase separation as defined by distance of H$_f$ to the tie line. Compounds with $dH$=0 are ground states. Compounds with 0$<dH<$20 meV/atom are termed metastable. A 20 meV/atom cutoff, a typical size of the \emph{relative} contribution to the Gibbs energy from the vibrational entropy term at $T\sim1000$ K~\cite{Kolmogorov2010}, is chosen to distinguish between possible metastable and unstable phases. Compounds with $dH$ values greater than 20 meV/atom are not included here.}
\label{tab:meta}
\end{longtable}

\section*{Acknowledgements}
A.N.K. acknowledges partial support from the EPSRC (CAF EP/G004072/1 ).

\bibliographystyle{myapsrev}

\pagebreak

\section{Supplemental Material}

\noindent{
In support of the main paper this supplementary materials contains the following descriptions and datasets:
\begin{itemize}
\item  A description of the magnetic ordering in elemental ground state structures for selected 3d transition metals;
\item  A list of compositions searched using the evolutionary algorithm for 0 GPa and 30 GPa (Table \ref{tab:evsearch});
\item  A description of the high-throughput density functional theory framework in MAISE;
\item  Full structural information of all the stable structures at 0 GPa (Table \ref{tab:struc0});
\item  Full structural information of select  metastable structures at 0 GPa (Table \ref{tab:strucMS0});
\item  Full structural information of all the stable structures at 30 GPa (Table \ref{tab:struc30}).
\end{itemize}

\begin{table}[h]
\caption{The compositions run with evolutionary search, excluding the Fe-B, Cr-B, and Ca-B systems studied in the author's  previous papers~\cite{Kolmogorov2012,Kolmogorov2010,Niu2012,shah2013}}\label{tab:evsearch}
\begin{tabular}{l  l l l l l l}
\hline\hline
Pressure:&\multicolumn{5}{ c }{Compositions}\\
       0 GPa: & BeB$_6$ &Cu$_4$B$_8$&Cu$_4$B$_4$&La$_8$B$_8$&LaB&Mg$_2$B$_{12}$\\
                   &Mg$_4$B$_{24}$&MgB$_{6}$&Mn$_2$B$_{4}$&Mn$_4$B$_{16}$&Mn$_2$B$_{8}$&Os$_8$B$_{8}$\\
                   &Os$_3$B$_{3}$&Os$_4$B$_{4}$&Pt$_2$B$_{2}$&Pt$_2$B$_{4}$&Pt$_4$B$_{4}$&Pt$_6$B$_{6}$\\
                   &Pt$_3$B$_{3}$&Pt$_6$B$_{4}$&Rh$_4$B$_{4}$&W$_2$B$_{6}$&W$_2$B$_{8}$&W$_3$B$_{12}$\\
                   &W$_4$B$_{12}$&W$_2$B$_{8}$&Zn$_4$B$_{4}$&Zn$_4$B$_{8}$&Hf$_8$B$_{8}$&Hf$_4$B$_{16}$\\
                   &Ir$_{3}$B$_{2}$&Ir$_{6}$B$_{4}$&Ir$_{9}$B$_{6}$&Ir$_{12}$B$_{8}$&K$_{4}$B$_{4}$&K$_{8}$B$_{8}$\\
                   &K$_{1}$B$_{1}$&Mo$_{4}$B$_{10}$&Mo$_{8}$B$_{20}$&Mo$_{2}$B$_{5}$&Mo$_{4}$B$_{12}$&MoB$_{3}$\\
                   &Mo$_{2}$B$_{6}$&Mo$_{2}$B$_{8}$&Mo$_{4}$B$_{16}$&MoB$_{4}$&Ni$_{4}$B$_{12}$&Ni$_{6}$B$_{18}$\\
                   &Ni$_{8}$B$_{24}$&Ni$_{2}$B$_{6}$&Ni$_{8}$B$_{8}$&\\
       30 GPa: & Ba$_4$B$_4$ &Ba$_4$B$_8$&Ba$_8$B$_4$&Be$_2$B$_6$&Be$_3$B$_6$&Be$_8$B$_8$\\
                     &Be$_2$B$_8$&Be$_4$B$_8$&K$_4$B$_4$&K$_2$B$_8$&K$_2$B$_{12}$&K$_4$B$_{12}$\\
                     &Li$_{6}$B$_{2}$&Li$_{8}$B$_{2}$&Li$_{16}$B$_{4}$&Li$_{2}$B$_{6}$&Li$_{2}$B$_{8}$&Li$_{4}$B$_{8}$\\
                     &Li$_{8}$B$_{8}$&Li$_{2}$B$_{12}$&Li$_{4}$B$_{12}$&Li$_{4}$B$_{16}$&Li$_{6}$B$_{16}$&Li$_{2}$B$_{2}$\\
                     &Li$_{3}$B$_{3}$&Na$_{4}$B$_{8}$&Na$_{2}$B$_{6}$&Na$_{2}$B$_{12}$&Sr$_{4}$B$_{4}$&Sr$_{4}$B$_{8}$\\
                     &Sr$_{2}$B$_{12}$&Sr$_{16}$B$_{4}$&Y$_{2}$B$_{8}$&Y$_{4}$B$_{16}$&Y$_{2}$B$_{8}$&Cs$_{4}$B$_{4}$\\
                     &Cs$_{4}$B$_{8}$&&&&&\\
\hline\hline
\end{tabular}
\end{table}

The chemical potentials of each metal at the respective pressure are determined for the calculation of the formation enthalpy (Eqn. (1)).  For the majority of these metals the lowest energy phase was either bcc, fcc, or hcp. However, the 3$d$ Cr, Mn, Fe, Co, and Ni metals require more detailed calculations to account for the ground state magnetic configurations. Cr is calculated to be antiferromagnetic (AFM) in a two-atom bcc unit cell. Fe is ferromagnetic bcc. Co is ferromagnetic hcp.  Ni is ferromagnetic fcc.  The known ground state for Mn is a large complex non-collinear distorted bcc structure~\cite{Hobbs2003a,Hafner2003b}.  However, due to the size and complexity of $\alpha$-Mn the energetically preferred structure of the bcc, fcc, and hcp structures was selected.  It is known that for GGA this correctly favors an AFM fcc structure, while LDA incorrectly favors AFM hcp~\cite{Hafner2003b,Stojic2008}.  Therefore, in this work we use the antiferromagnetic fcc structure for both GGA and LDA (forced). We have checked that the 215 meV/atom difference for the LDA does not affect the stability of the Mn-rich compounds.

The high-throughput density functional theory framework (implemented under the auspices of MAISE) used to study the structures for each system performed three geometry optimizations with progressingly strict convergent criteria to allow structures determined for other systems to converge to realistic cell parameters and atomic positions.  Each optimization converged to a strain below 2-3 kbar (0.2 - 0.3 GPa) with interatomic forces below 0.01 eV/$\AA$.  Spot checks were performed to determine that the electronic steps converged appropriately with the convergence criteria  changed accordingly  (usually increased) to converge the structure.



\bibliographystyle{apsrev}

\end{document}